\documentclass[12pt,aps,pre,preprint]{revtex4}

\usepackage{amssymb}
\usepackage{amsmath}
\usepackage[dvips]{graphicx}
\usepackage{colordvi}
\usepackage{times}

\newcommand{\bea}{\begin{eqnarray}}
\newcommand{\eea}{\end{eqnarray}}
\newcommand{\be}{\begin{equation}}
\newcommand{\ee}{\end{equation}}

\begin{document}

\title{Current in a quantum driven thermostatted system with off-diagonal disorder}

\author{Matteo Colangeli \email{matteo.colangeli@polito.it}}
\affiliation{Dipartimento di Matematica, Politecnico di Torino, Corso Duca degli Abruzzi
24, I-10129 Torino, Italy}

\author{Marco Pizzi}
\affiliation{Eltek S.p.A., Strada Valenza 5/A, 15033 Casale Monferrato, Italy}

\author{Lamberto Rondoni}
\affiliation{Dipartimento di Matematica, Politecnico di Torino, Corso Duca degli Abruzzi
24, I-10129 Torino, Italy.\\
INFN, Sezione di Torino, Via P. Giura 1, I-10125, Torino, Italy}


\begin{abstract}
We analyze a one-dimensional quantum model with off-diagonal disorder, consisting of a sequence of potential energy barriers whose width is a random variable either uniformly or ``half-normally'' distributed, subjected to an external electric field. We shed light on how the microscopic disorder affects the value of the transmission coefficient, and on the structure of the fluctuations around the solutions corresponding to the regular lattice configuration. We also characterize the asymptotic limit obtained by letting the number of barriers diverge, while their total width is kept constant. Thus, we explain the novelty of our method with respect to the standard thermodynamic limit discussed in the literature, and also evidence the onset of a large deviations principle for the transmission coefficient.
\end{abstract}

\maketitle

\section{Introduction}
\label{sec:sec0}

Nonequilibrium thermodynamics is based on the notion of space and time scales separation and on the assumption of \textit{local equilibrium} \cite{dgm,Liboff,GibRon,matt2,matt3,matt4}. The theory of large deviations, in particular, helped to understand and interpret the role of the fluctuations in nonequilibrium systems \cite{touch,maes,colirr,BPRV}. On the other hand, recent technological advances on the nanoscale science and technology demand an extension of the theoretical apparatus and foster a statistical mechanical approach to systems of relatively small numbers of degrees of freedom. In such systems the microscopic, mesoscopic and macroscopic scales can not be sharply separated, and the physical properties of microscopic devices widely fluctuate with respect to their mean values, violating the standard thermodynamic laws which describe macroscopic fields. In this work we face these issues by considering a variant of the original Anderson model, which is the prototype of a disordered solid \cite{Vulp}.
In particular, we investigate the role of the microscopic disorder on the transmission coefficient of one-dimensional systems consisting of a sequence of $N$ barriers, with random widths, and $N-1$ wells, under the constraint that the sum of the barrier widths and the total length of the system are fixed and do not change with $N$. We then introduce a classical thermostat at given temperature $T$ and an external electric potential $V_\ell-V_r$.
Furthermore, we do not introduce simplifying assumptions such as the ``tight-binding'' approximation introduced by Anderson in his pioneering paper \cite{Ander} on localization effects in disordered solids. Therefore, our model enjoys a purely off-diagonal disorder \cite{TC,SE, Izrailev} which concerns only the tunneling couplings among the wells, leaving unaffected the energies of the bound states within the wells. This is not the case of the original tight-binding model, whose random fluctuations only concern the energy of a bound state. In turn, while in Anderson's model increasing the number of barriers corresponds to taking the large system limit, in our case it corresponds to distribute more finely the same amount of insulating material within the fixed length of the system.\\ 
The introduction of an external field allows to extend to ``nonequilibrium'' the results previously obtained in the analysis of the model treated in  \cite{ColRon}, which are recovered, as shown below, in the limit of vanishing external fields. We, thus, investigated the effect induced by this kind of disorder at the mesoscopic scale on the transmission coefficient and we shed light on the structure of its fluctuations.
Our results can be summarized as follows:
\begin{itemize}
\item There are no localization effects for the equilibrium distribution of energies at temperature $T = 300 K$:
positive currents persist even in the large $N$ limit. 
\item Furstenberg type theorems \cite{Vulp} do not apply. The reason is that the product of the random matrices yielding the transmission
coefficient for a given choice of $N$ barriers changes, in order to preserve the length of the system and 
the sum of the barrier widths, when the $N+1$-th barrier is introduced. 
\item The value of the transmission coefficient, averaged over an ensemble of disordered configurations, 
is close, for large $N$, to the value corresponding to the ordered sequence of equally spaced barriers 
and wells, which is bounded away from zero.
\item There is a scale for $N$, above which the (always positive) transmission coefficient does not depend 
on the specific realization of the disorder, but still depends on $N$, and there is another scale above which even the dependence on $N$ is eliminated. We call ``mesoscopic'' the first, and ``macroscopic'' the latter scale, since it represents macroscopic 
nanostructured materials. This means that all realizations of the disorder become equivalent in the $N\to\infty$ limit.
\item At room temperature, the probability distribution function (PDF) of the time independent transmission
coefficients of the different realizations of the system satisfies a principle of large deviations. Furthermore, the peak of this PDF 
corresponds to the transmission coefficient of the regular realizations.
\item Our $N\rightarrow \infty$ limit, representing a macroscopic object at given temperature, which is microscopically randomly structured, leads to radically different results from the usual macroscopic limits. In particular, it leads, in certain systems, to the experimentally verifiable lack of localization. This is relevant in situations complementary to those described by the standard theories.

\end{itemize}

\section{The model}
\label{sec:sec1}

\begin{figure}
   \centering
   \includegraphics[width=0.8\textwidth]{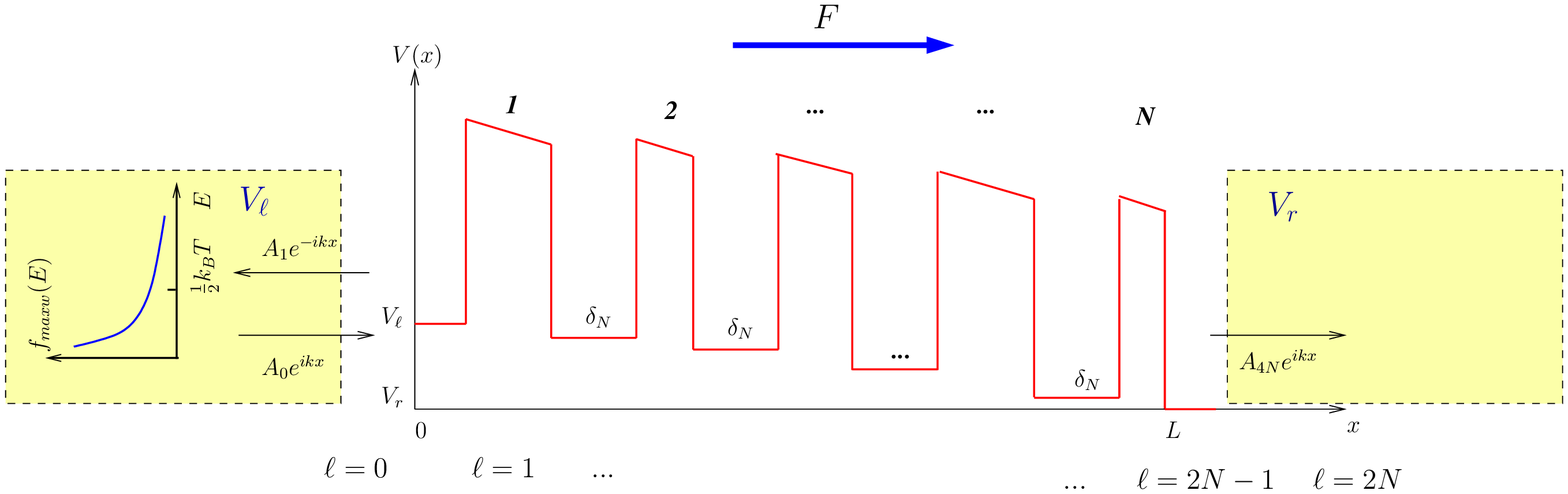}
   \caption{(Color online) $1$-D multiple-barrier system, consisting in a sequence of $2N+1$ regions: $N$ barriers, whose width is uniformly random distributed, $(N-1)$ conducting regions of constant width $\delta_N$, and the two boundary regions (shadowed areas), on the left and on the right, representing the electrodes. The latter are modelized as a classical thermostat at temperature $T = 300 K$ and electric potential $V_\ell$ (electrode on the left) and $V_r$ (electrode on the right). The mean energy of the plane waves entering from the left boundary is given by $\frac{1}{2}k_B T$. The picture illustrates the case characterized by vanishing electric field in the conducting regions.}\label{barriers}
\end{figure}

Our one-dimensional model of a macroscopic semiconductor device consists of an array of $N$ potential barriers and $N-1$ conducting regions (wells), in contact with one electrode which acts as an external thermostat at temperature $T=300 K$. The particles leaving this thermostat are subjected to an external electric field $F$, cf. Fig. \ref{barriers}. The barriers have a constant height $V(x)=V$ while their width is either uniformly or ``half-normally'' randomly distributed. For any $N$, the widths of the conducting regions take a constant value $\delta_N$.
We denote by $L$ the fixed total length of the sample, by $L^{is}$ the fixed sum of the widths of all the barriers (i.e. the total length of the insulating region), and by $\beta$ the fixed ratio between insulating and conducting lengths, so that 
\be 
L=(1+\beta)(N-1)\delta_N  \label{L}
\ee
holds. 
To compute the current, we study the steady state Schr\"{o}dinger Equation (SE):

\be
-\frac{\hbar^2}{2m}\frac{d^2}{dx^2}\psi(x)+(V-e F x)\psi(x)=E\psi(x), \quad x\in[0,L] \label{se}
\ee
where $m$ is the mass of the particle, $e$ the electronic charge and $F$ the magnitude of the external electric field which takes the values $F^{is}$ inside the barriers and $F^{con}$ in the conducting regions. Due to the electric field, the potential energy decreases monotonically from $V_\ell$, on the left boundary, down to $V_r$, on the right boundary, with a slope given by, respectively, $-e F_{is}$ within the barriers and $-e F_{c}$ in the conducting regions.
Let us also introduce the parameter $r=F_{is}/F_c < \infty$, which allows to consider the presence of a nonvanishing electric field even within the wells.
Therefore, the energy of the electric field acting on the system, denoted by $E_v$, amounts to
\be
E_v= e (V_\ell-V_r)= e F_{c} L_{c} (r \beta + 1) \label{Ev}
\ee
The boundary conditions prescribe $A_0>0$ for the amplitude of the plane wave entering from the left boundary and $A_{4N+1}=0$ (no wave enters or is reflected from the right boundary).
The barriers are delimited by a set of $2N$ points, denoted by $x_0=0,...,x_{2N-1}=L$ in Fig. \ref{barriers}, hereafter called \textit{nodes} of discontinuity of the potential. The left boundary consists of a classical thermostat at temperature $T$, from which particles emerge at node in $x_0$ as plane waves, with energies distributed according to the Maxwell-Boltzmann distribution. Differently, no particles come from the electrode on the right. 

Thus, denoting by $\mathcal{U}_\ell$ the $\ell$-th region, for $\ell\in\{0,2,...,2N\}$, the solutions of eq. (\ref{se}) take the form:

\be
\psi_\ell(x)=\begin{array}{cc}

A_{2\ell} Ai_\ell(x)+A_{2\ell+1} Bi_\ell(x), & \quad \text{if $x\in \mathcal{U}_\ell$} 
               \end{array}   \label{psi}
\ee
where $Ai_\ell(x)$ and $Bi_\ell(x)$ denote the Airy functions.
In each of the conducting regions, one may define the steady state \textit{currents} as:
\bea
j_\ell(x)&=&\frac{\hbar}{2mi}\left[\psi_\ell(x)^*\left(\frac{d}{dx}\psi_\ell(x)\right)-(\frac{d}{dx}\psi_\ell(x)^*)\psi_\ell(x)\right]=\nonumber\\
&=&j_\ell^{tr}(A_{2\ell})-j_\ell^{rif}(A_{2\ell+1}) \quad , \label{curr}
\eea
where the $^*$ denotes complex conjugation, $j_\ell^{tr}(A_{2\ell})=\hbar k/m|A_{2\ell}|^2$ denotes the current transmitted from the $(\ell-1)$-th barrier on the left (or, for $\ell=0$, from the thermostat located at the left boundary) and $j_\ell^{tr}(A_{2\ell+1})=\hbar k/m|A_{2\ell+1}|^2$ denotes the current reflected from the $(\ell+1)$-th barrier, cf. Fig. \ref{barriers}. Then, the application of the BenDaniel-Duke boundary conditions \cite{harris}, which require the continuity of $\psi_\ell(x)$ and $\frac{d}{dx}\psi_\ell(x)$ at the nodes, results in the constancy of the value $j_\ell(x)$ across the wells and entails $j_\ell(x)=j_{2N}(x)$, for every even $\ell$. Equation (\ref{curr}), together with Eq. (\ref{psi}), leads to the following definition of the transmission coefficient $J$ across the system:
\be
J=\frac{j_{2N}^{tr}(A_{4N})}{j_0^{tr}(A_{0})}=\frac{|A_{4N} |^2}{|A_{0}|^2} \label{J}
\ee
which depends on the several parameters of the model, such as the number of barriers $N$, the energy $E_v$ of the applied electric field, the ratio $r$ and the disordered configuration of the sequence of barriers. In Sec. \ref{sec:sec4} we focus, in particular, on the structure of the fluctuations of $J$ as a function of the realization of the disorder. We also discuss how the magnitude of these fluctuations depends on $N$, for a given distribution of noise realizations, by exploring a large range of scales: from the microscopic one, where $N=O(1)$, up to the macroscopic one, with $N\gg 1$.
In order to numerically compute the coefficient $J$ as a function of the various parameters of the model, it proves convenient to rescale Eq. (\ref{se}) with respect to characteristic quantities, in order to rewrite it in a dimensionless form. For this purpose, let us introduce $x=L \tilde{x}$, with $L$ given by (\ref{L}), $E=\tilde{E}E_T$, $\psi(x)=\tilde{\psi}(\tilde{x})(\sqrt{L})^{-1}$, $V=\tilde{V}E_T$, $F_c=\tilde{F}_c E_T/(e L)$, with $E_T=k_B T$ (i.e. twice the mean kinetic energy of the plane waves entering the bulk from the left side). Moreover, by introducing the scalar parameter $\alpha=\hbar^2/(2 m L^2 E_T)$, one obtains the following expression for the dimensionless wavevectors: $\tilde{k}=\sqrt{\alpha}\sqrt{\tilde{E}}$ and $\tilde{z}=\sqrt{\alpha}\sqrt{\tilde{V}-\tilde{E}}$. In the sequel we will refer to the dimensionless quantities and, to this aim, we may omit the tilde symbols, for sake of simplicity.
The dimensionless version of eq. (\ref{se}), then, attains the form:
\be
- \alpha \frac{d^2}{dx^2}\psi(x)+(V(x)- e F x)\psi(x)=E\psi(x), \quad x\in[0,1] \label{se2}
\ee
and it represents the SE which will be solved numerically with the aforementioned conditions at the nodes.

\section{The transfer matrix technique}
\label{sec:sec2}

Let us now describe our method of solution of the SE, eq. (\ref{se2}), which follows Refs. \cite{Vulp,harris} and is referred to as the Transfer Matrix (TM) technique.
Using eqs. (\ref{psi}), the BenDaniel-Duke boundary conditions on the generic $\ell$-th node, with $\ell\in\{0,1,...,2N-1\}$, cf. Fig. \ref{barriers}, read as:
\be
\left\{\begin{array}{c}
\psi_\ell(x_\ell)=\psi_{\ell+1}(x_\ell) \\
\psi_\ell'(x_\ell)=\psi_{\ell+1}'(x_\ell)
\end{array} \right. \quad , \label{BD}
\ee
where $x_\ell(x)=\sum_{i=1}^{\ell/2}\tilde{\lambda}_i+\delta \frac{\ell}{2}$ if $\ell$ is even and $x_\ell(x)=\sum_{i=1}^{(\ell+1)/2}\tilde{\lambda}_i+\delta \frac{\ell-1}{2}$ if $\ell$ is odd and where $\tilde{\lambda}_i$ denotes the random width of the $i$-th barrier.
Thus, in matrix form, eqs. (\ref{BD}) takes the form:
\bea
\mathbf{M}_0(x_0) \cdot \left( \begin{array}{c}
A_0 \\
A_1 
\end{array} \right)&=&\mathbf{M}_1(x_0)\cdot\left( \begin{array}{c}
A_2 \\
A_3 
\end{array} \right) \nonumber\\
\mathbf{M}_2(x_1) \cdot\left( \begin{array}{c}
A_2 \\
A_3 
\end{array} \right)&=&\mathbf{M}_3\cdot(x_1)\left( \begin{array}{c}
A_4 \\
A_5 
\end{array} \right) \nonumber\\
\mathbf{M}_4(x_2) \cdot\left( \begin{array}{c}
A_4 \\
A_5 
\end{array} \right)&=&\mathbf{M}_5(x_2)\cdot\left( \begin{array}{c}
A_6 \\
A_7 
\end{array} \right) \label{trmat}\\
&\vdots&\nonumber\\
\mathbf{M}_{4N-2}(x_{2N-1}) \cdot\left( \begin{array}{c}
A_{4N-2} \\
A_{4N-1} 
\end{array} \right)&=&\mathbf{M}_{4N-1}(x_{2N-1})\cdot\left( \begin{array}{c}
A_{4N} \\
A_{4N+1} 
\end{array} \right) \nonumber \quad ,
\eea
where the $2\times 2$ matrices of coefficients $\mathbf{M}_{2\ell}(x_\ell)$ and $\mathbf{M}_{2\ell+1}(x_\ell)$ read
\be
\mathbf{M}_{2\ell}(x_{\ell})=\left( \begin{array}{c c}
Ai_\ell(x_\ell) & Bi_\ell(x_\ell) \\
Ai_\ell^{'}(x_\ell) & Bi_\ell^{'}(x_\ell) 
\end{array} \right)
\ee and 
\be
\mathbf{M}_{2\ell+1}(x_\ell)=\left( \begin{array}{c c}
Ai_{\ell+1}(x_\ell) & Bi_{\ell+1}(x_\ell) \\
Ai_{\ell+1}^{'}(x_\ell) & Bi_{\ell+1}^{'}(x_\ell)
\end{array} \right)
\ee
Assuming that the amplitude $A_0$ of the incoming wave $\psi_0$ is known and that $A_{4N+1}=0$ because there is no reflection from the right boundary in the $2N$-th region, Fig. \ref{barriers}, then the linear system (\ref{trmat}) corresponds to a set of $4N$ equations in the $4N$ variables $\{A_1,...,A_{4N}\}$.
Skipping the spatial dependence of the matrices to keep our notation simple, eq. (\ref{trmat}) implies
\be
\left( \begin{array}{c}
A_{0} \\
A_{1} 
\end{array} \right)=\underbrace{\mathbf{M}_{0}^{-1}\cdot\mathbf{M}_{1}\cdot\mathbf{M}_{2}^{-1}\cdot\mathbf{M}_{3}\cdot...\cdot\mathbf{M}_{4N-2}^{-1}\cdot\mathbf{M}_{4N-1}}_{\mathbf{M}}\cdot \left( \begin{array}{c}
A_{4N} \\
0 
\end{array}\right)= \mathbf{M}\cdot\left( \begin{array}{c}
A_{4N} \\
0 
\end{array}\right)\label{trmat2} \quad ,
\ee
where we have defined the $2\times 2$ matrix $\mathbf{M}$.
Eq. (\ref{trmat2}) leads to 
\be
A_{0}=M_{11} A_{4N} \label{trmat3}
\ee
with $M_{11}$ denoting the element on the top left corner of the matrix $\mathbf{M}$. Therefore, the relation (\ref{trmat3}) allows us to rewrite eq. (\ref{J}) in the form:
\be
J=\frac{A_{4N}^*A_{4N}}{A_{0}^*A_{0}}=\frac{1}{|M_{11}|^2} \label{J2}
\ee
Our results will be illustrated in Sec. \ref{sec:sec3} below.

\section{Microscopic disorder and the external thermostat}
\label{sec:sec3}

We characterize the presence of disorder in the sequence of barriers letting $\rho(\hat{\lambda}) d\hat{\lambda}$ be the probability distribution of the widths $\hat{\lambda}_i$ of a generic barrier, with $i =\{1,...,N\}$, to take values in a range $d\hat{\lambda}$ centered on $\hat{\lambda}$. 
In particular, we used pseudo-random generators to investigate two relevant choices for $\rho(\hat{\lambda})$. The first is the \textit{uniform} density, with support on the unit interval, while the other is the density $\rho=\sqrt{\frac{2}{\pi}}e^{-\frac{\hat{\lambda}^2}{2}}$, supported on $\mathbb{R}^+$, which is obtained from the gaussian density $\rho=\mathcal{N}(0,1)$ by retaining only the positive values of the $\hat{\lambda}_i$'s. Each of the two distributions is characterized by the corresponding mean $\langle \hat{\lambda}\rangle$ and variance $\hat{\sigma}^2$ \footnote{For a uniform density $\langle \hat{\lambda}\rangle =0.5$, $\hat{\sigma}^2 = 1/12$, whereas for the ``half-normal'' density defined above, $\langle \hat{\lambda}\rangle =2/\pi$, $\hat{\sigma}^2 = 1-2/\pi$.}. It proves useful to introduce, for both these distributions, the \textit{realization} mean and variance, defined, 
respectively, as
\bea
\hat{\lambda}_B&=& \frac{1}{N}\sum_{i=1}^{N}\hat{\lambda}_i ,\nonumber \\
\hat{W}_N^2&=& \frac{1}{N}\sum_{i=1}^{N}(\hat{\lambda}_i-\langle \hat{\lambda}\rangle)^2 . \nonumber
\eea
In the large $N$ limit, the random variable $\hat{\lambda}_B$, which varies from realization to eralization fo the sequence of barriers, converges in probability to the mean $\langle \hat{\lambda}\rangle$, while the random variable $\hat{W}_N^2$ converges with probability $1$ to $\hat{\sigma}^2$. 
Since we use dimensionless variables in eq. (\ref{se2}), we introduce the rescaled barrier width as:
\be
\lambda_i=c \hat{\lambda}_i, \quad  \hbox{with} \quad c=\frac{\beta}{(1+\beta) N \hat{\lambda}_B} \quad . \label{constr}
\ee 
Therefore, for any given $N$ and $\beta$, the rescaled mean
$$
\lambda_B=\frac{\beta}{(1+\beta)N}
$$ 
is no longer a random variable, and attains the same constant value independently of the realization, hence on the density $\rho$. On the other hand, the rescaled realization variance 
$$
W_N^2=\frac{c^2}{N}\sum_{i=1}^{N}(\hat{\lambda}_i-\hat{\lambda}_B)^2
$$
remains a random variable which, for large $N$ converges to $\sigma^2=c^2 \hat{\sigma}^2$ with probability $1$. We introduce the vector-valued random variable $\Lambda_N$, defined by 
\be
\Lambda_{N}=\left\{\lambda_1,...,\lambda_N \right\} \quad , \label{Lambda}
\ee 
which corresponds to a given realization of the sequence of barriers and will be referred to as a \textit{microscopic configuration}.
For given $\beta$ and $N$, one may, then, consider the collection $\Omega=\{\Lambda_{N}^{(1)},...,\Lambda_{N}^{(N_r)}\}$ of $N_r$ random realizations of the sequence of barriers which have been constructed numerically. 
Then, the average of a random observable $\mathcal{O}$ over the sample $\Omega$  simply reads as: 
\be
\langle \mathcal{O} \rangle_{\Omega}=\frac{1}{N_r}\sum_{\mu=1}^{N_r}\mathcal{O}\left(\Lambda_{N}^{(\mu)}\right) \label{ranav}
\ee
Among the possible configurations, the regular one
\be
\Lambda_{N}^{B}=\left\{\lambda^B,...,\lambda^B \right\} \label{Bloch} \quad,
\ee
which approximates the \textit{infinite superlattice} of the literature on Bloch waves \cite{harris,mermin}, will be crucial also in our work.\\
In our numerical simulations we investigated the behavior of the coefficient $J(N,\Lambda_N,V,E;T)$ (we do not explicitly indicate the dependence on the parameters $r$, $\beta$ and $E_v$, not to overload the notation) at a given temperature $T$ on the parameters of the model, in particular the number of barriers $N$, the height of the barrier $V$, and, mostly, the microscopic configuration $\Lambda_N$. 

\begin{figure}
\centering
\includegraphics[width=0.48\textwidth]{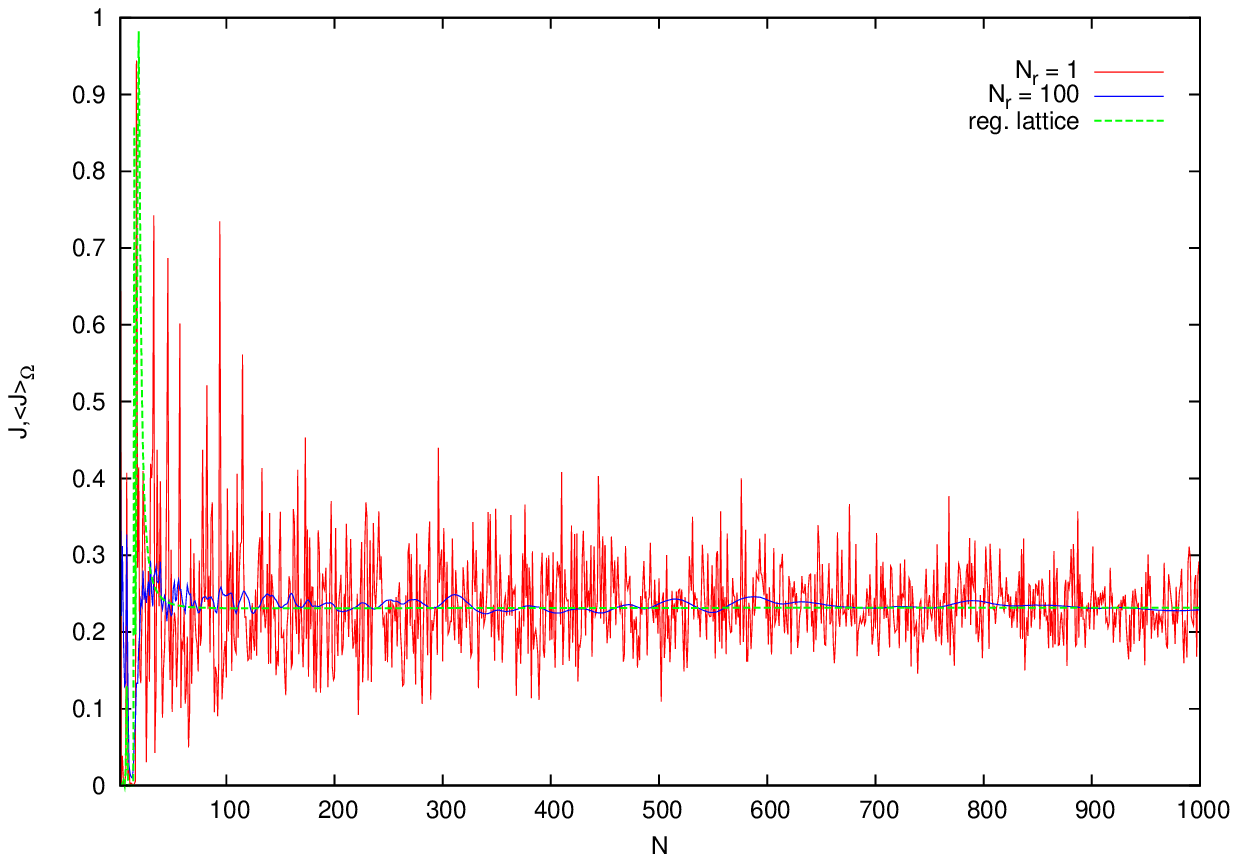}
\hspace{1mm}
\includegraphics[width=0.48\textwidth]{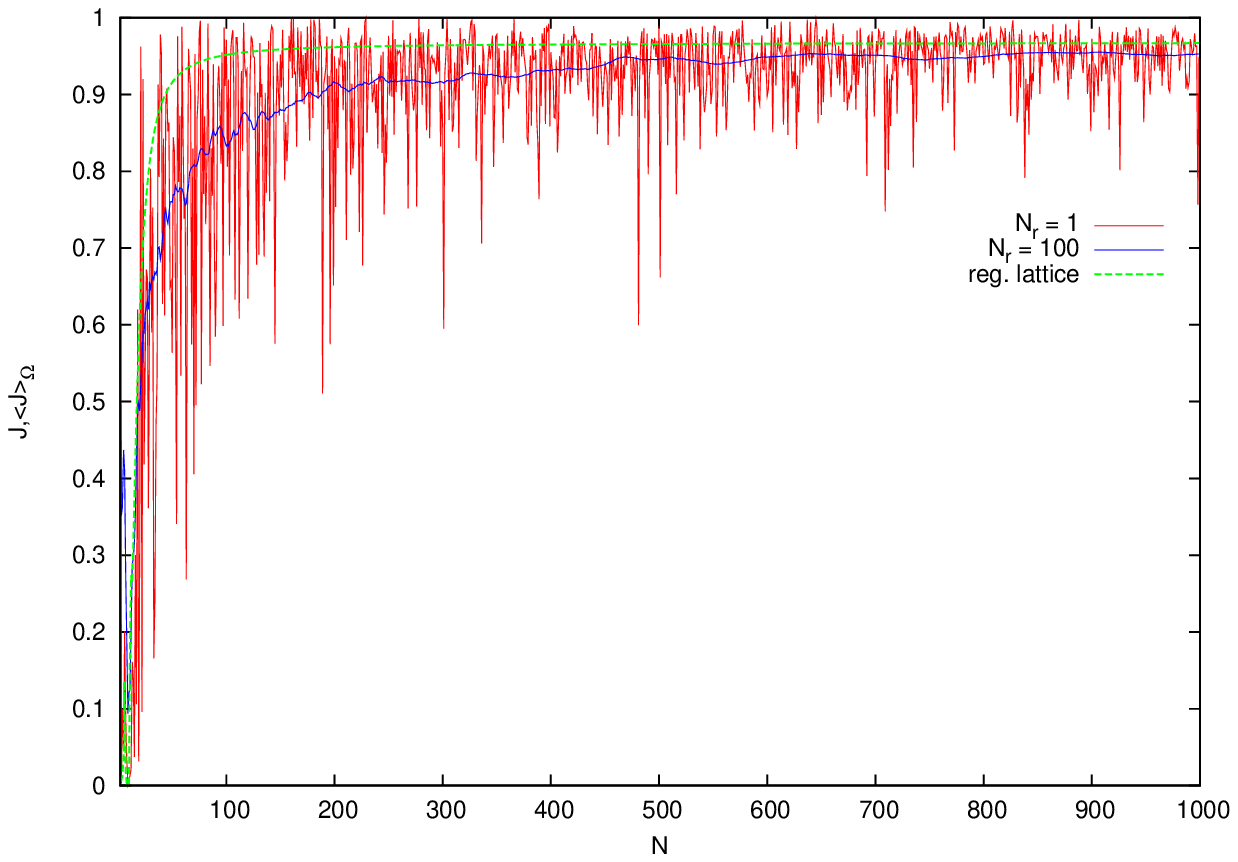}
\caption{(Color online) \textit{Left panel}: Behavior of $J(N,\Lambda_N,V,E;T)$ vs. $N$, for $V=20$, $E=9.60$, $E_v=1 eV$, $r=10^3$, $\beta=0.1$ and for different microscopic configurations: the red curve corresponds to a single uniformly distributed configuration $\Lambda_N$, the blue curve represents the average $\langle J\rangle_\Omega$ over an ensemble of uniformly distributed configurations, whereas the dashed green curve corresponds to the regular lattice configuration (\ref{Bloch}).
\textit{Right panel}: Same curves as those shown in the left panel, but here evaluated at $E=5.16$, where a condition close to the tunneling resonance is realized.}
\label{NEqmesk}
\end{figure}

Figure \ref{NEqmesk} illustrates the behavior of $J(N,\Lambda_N,V,E;T)$, $J(N,\Lambda_{N}^{B},V,E;T)$ and of the average $\langle J\rangle_\Omega$, at $T=300 K$, over an ensemble characterized by $\rho(\tilde{\lambda})=1$, for two different values of the energy $E$, one of which leads to a condition close to that of resonant tunneling.
The plots show the behavior of $J$ corresponding to a single disordered realization, of the average $\langle J\rangle_\Omega$, computed over a set of $N_r=10^2$ disordered realizations, and of the value, hereafter denoted as $J_B(N,V,E;T)$, corresponding to the regular lattice configurations (\ref{Bloch}). They also reveal that $\langle J\rangle_\Omega$ tends, for growing $N$, to the most probable value of $J(N,\Lambda_N,V,E;T)$, given by $J_B(N,V,E;T)$, as also discussed below. This holds for all values of $V$ and of $E$, and was already observed in the \textit{equilibrium} version of this model, discussed in \cite{ColRon}. Let us now consider a thermostat located at the left boundary, so that the plane waves entering the bulk have an energy obeying a classical equilibrium distribution at a given temperature $T$.
In the following plots we consider a one-dimensional Maxwellian probability density
\be
f_{eq}(E)=\sqrt{1/(\pi E)} e^{-E} \nonumber
\ee 
and we average over all energies to obtain 
\be
 J(N,\Lambda_N,V;T)=\int_{0}^{\infty} J(N,\Lambda_N,V,E;T) f_{eq}(E) dE \label{eqaver} 
\ee
where the coefficient $J$, defined in  eq. (\ref{J2}), is integrated in the r.h.s. of eq. (\ref{eqaver}).

\section{Nonequilibrium fluctuations}
\label{sec:sec4}

\begin{figure}
   \centering
   \includegraphics[width=0.48\textwidth]{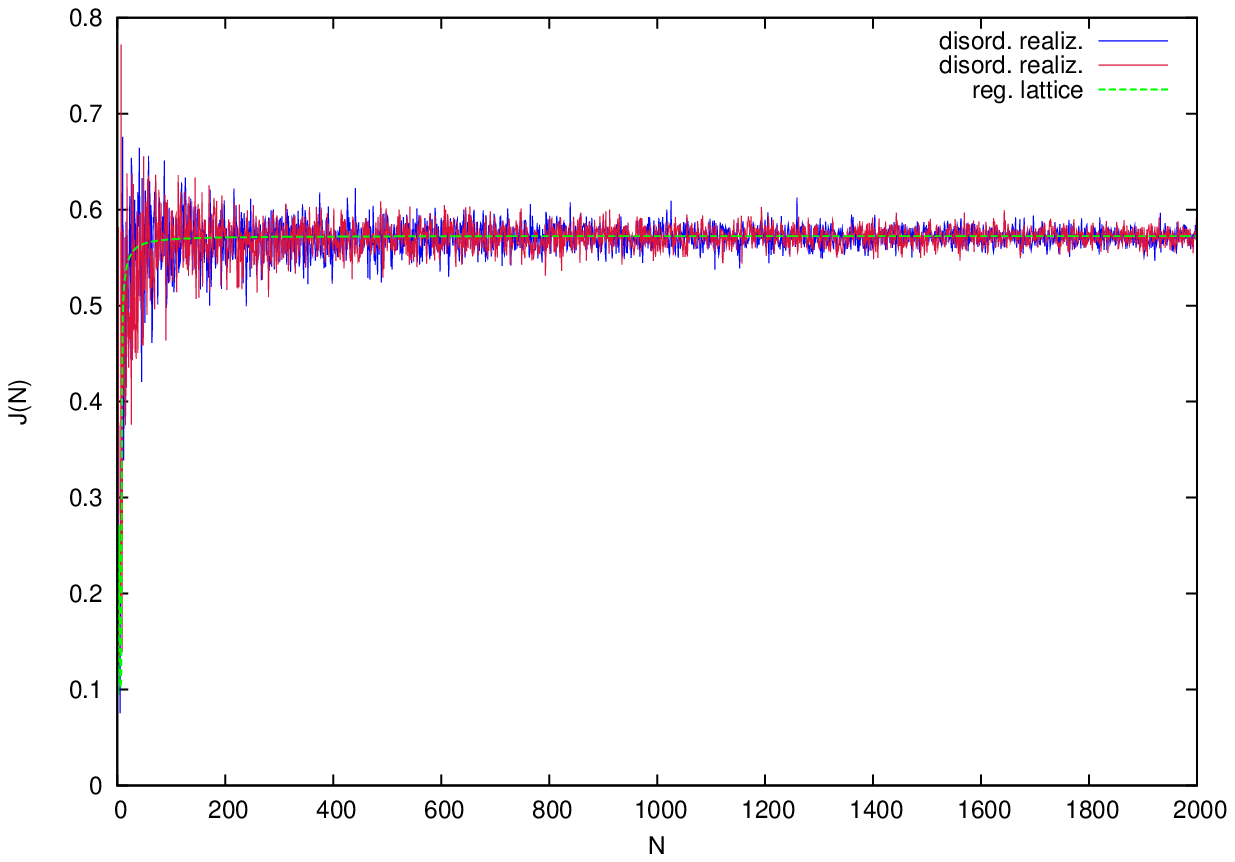}
   \hspace{1mm}
\includegraphics[width=0.48\textwidth]{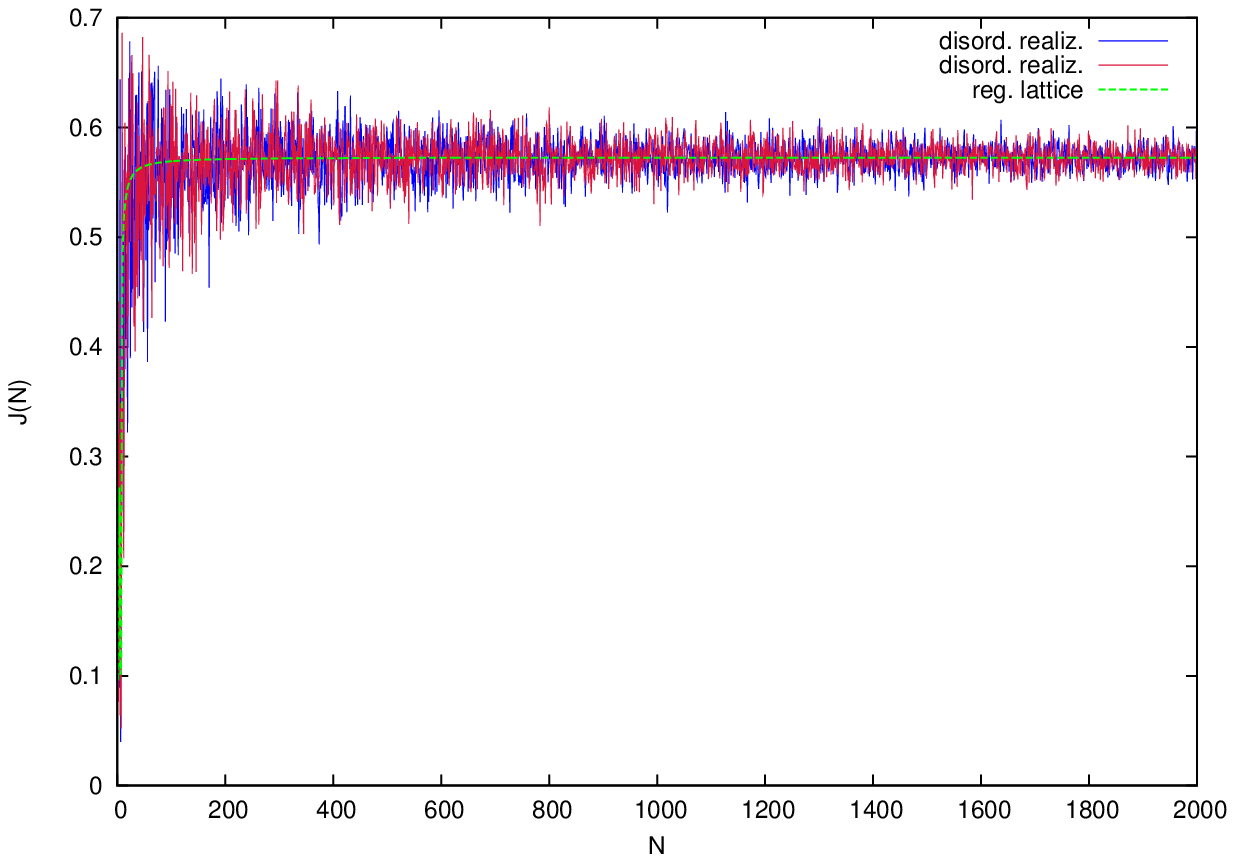}
   \caption{(Color online) \textit{Left panel}: Behavior of $J(N,\Lambda_N,V;T)$ vs. $N$ for $V=20$, $T=300 K$, $E_v=1 eV$, $r=10^3$, corresponding to two disordered configurations drawn from the ``half-normal'' distribution (dark and light blue curves) and to the regular lattice configuration (thick green dashed curve). \textit{Right panel}: Same as in the left panel, with two configurations obtained from the uniform probability density.}\label{noise1}
\end{figure}

We show here some numerical results concerning the value of $J$ for a single realization, averaged over the equilibrium distribution of energies, as shown in eq. (\ref{eqaver}). 
The typical behavior of $J(N,\Lambda_N,V;T)$ as a function of the number of barriers $N$ is illustrated in Fig. \ref{noise1}: the left panel shows the fluctuations, around the value $J_B(N,V,E;T)$, of the transmission coefficient pertaining to two disordered configurations obtained from the ``half-normal'' distribution introduced in sec. \ref{sec:sec3}. Similarly, the right panel displays the fluctuating behavior of $J$ with $N$ for two random uniformly distributed configurations.

Figure \ref{noise1} anticipates two further crucial aspects which will be addressed in more detail below. The first concerns the magnitude of the fluctuations of the values of $J(N,\Lambda_N,V;T)$, $\sigma_\rho(N,V)=\sqrt{\langle (J-\langle J\rangle_\Omega)^2\rangle_\Omega}$, which decreases with $N$. \\
This decay of the size of the fluctuations allows us to identify a ``mesoscopic'' scale $N_{meso}$. The number $N_{meso}$ depends on $V$ and the coefficient $J(N,\Lambda_N,V;T)$ depends only weakly on the microscopic configuration, if $N>N_{meso}$.
A second scale $N_{macro}>N_{meso}$ is identified such that $J(N,\Lambda_N,V;T)$ depends neither on the configuration nor on the number of barriers if $N> N_{macro}$. In the case $N> N_{macro}$ we practically have an infinitely finely structured sample of macroscopic fixed length. 

\begin{figure}
   \centering 
   \includegraphics[width=0.48\textwidth]{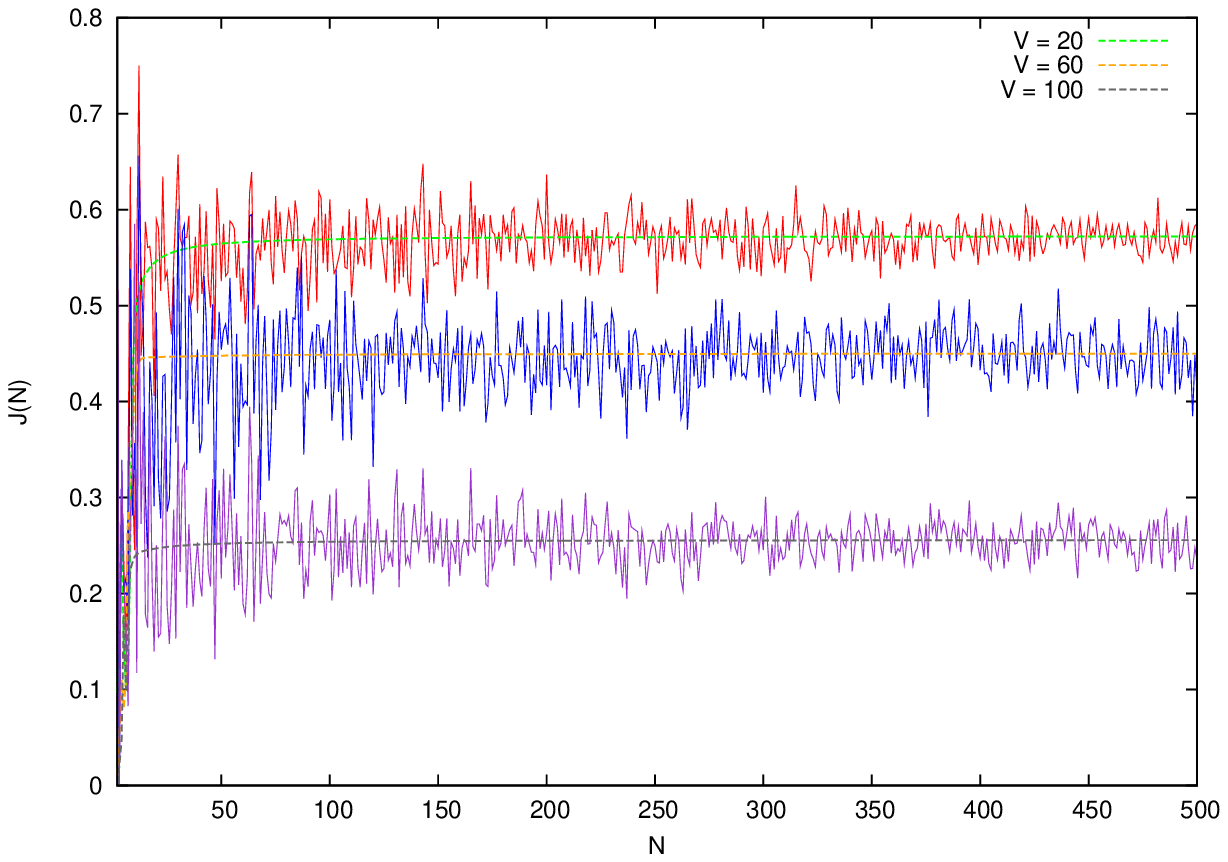}
   \hspace{1mm}
   \includegraphics[width=0.48\textwidth]{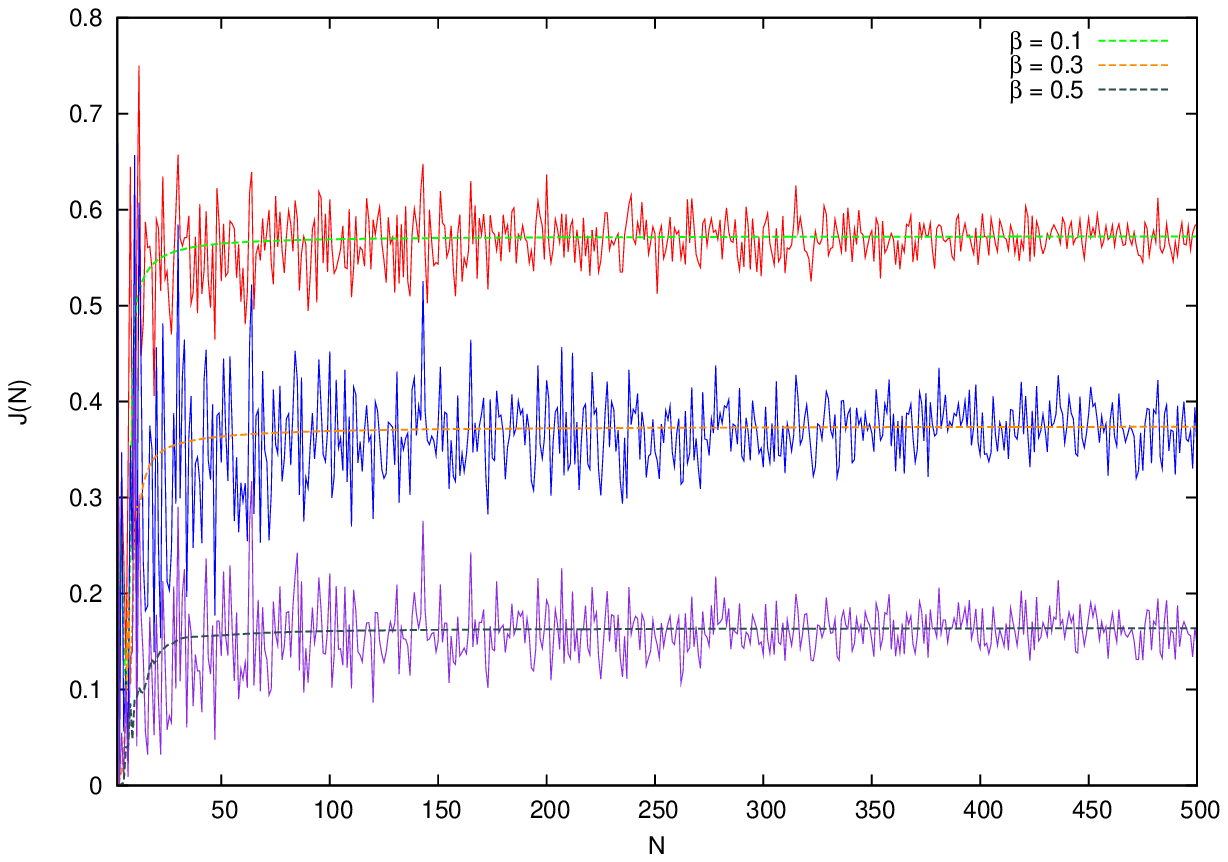}
   \caption{(Color online) \textit{Left panel}: Behavior of $J(N,\Lambda_N,V;T)$ vs. $N$ for $V=20$, $T=300 K$, $E_v=1 eV$, $r=10^3$ in an ensemble of random uniform configurations for different values of $V$. \textit{Right panel}: Behavior of $J(N,\Lambda_N,V;T)$ vs. $N$ for $V=20$, $T=300 K$, $E_v=1 eV$, $r=10^3$ in an ensemble of random uniform configurations for different values of $\beta$. }\label{noise2}
\end{figure}

We also investigated the dependence of $J(N,\Lambda_N,V;T)$ on $V$ and $\beta$. The left panel of Fig. \ref{noise2} corroborates, in the limit of large $N$, the numerical results illustrated in Fig. \ref{noise1}. Namely, the trend of the random values $J(N,\Lambda_N,V;T)$ to approach $J_B$ persists even when the potential energy of the barrier $V$, as well as the amount of insulating fraction in the system (related to $\beta$) are changed. This stems as one of the prominent features of our model. In particular, Fig. \ref{noise2} shows that an increase in $V$ or in $\beta$ produces a decrease of the observed transmission coefficient.

\begin{figure}
   \centering 
   \includegraphics[width=0.48\textwidth]{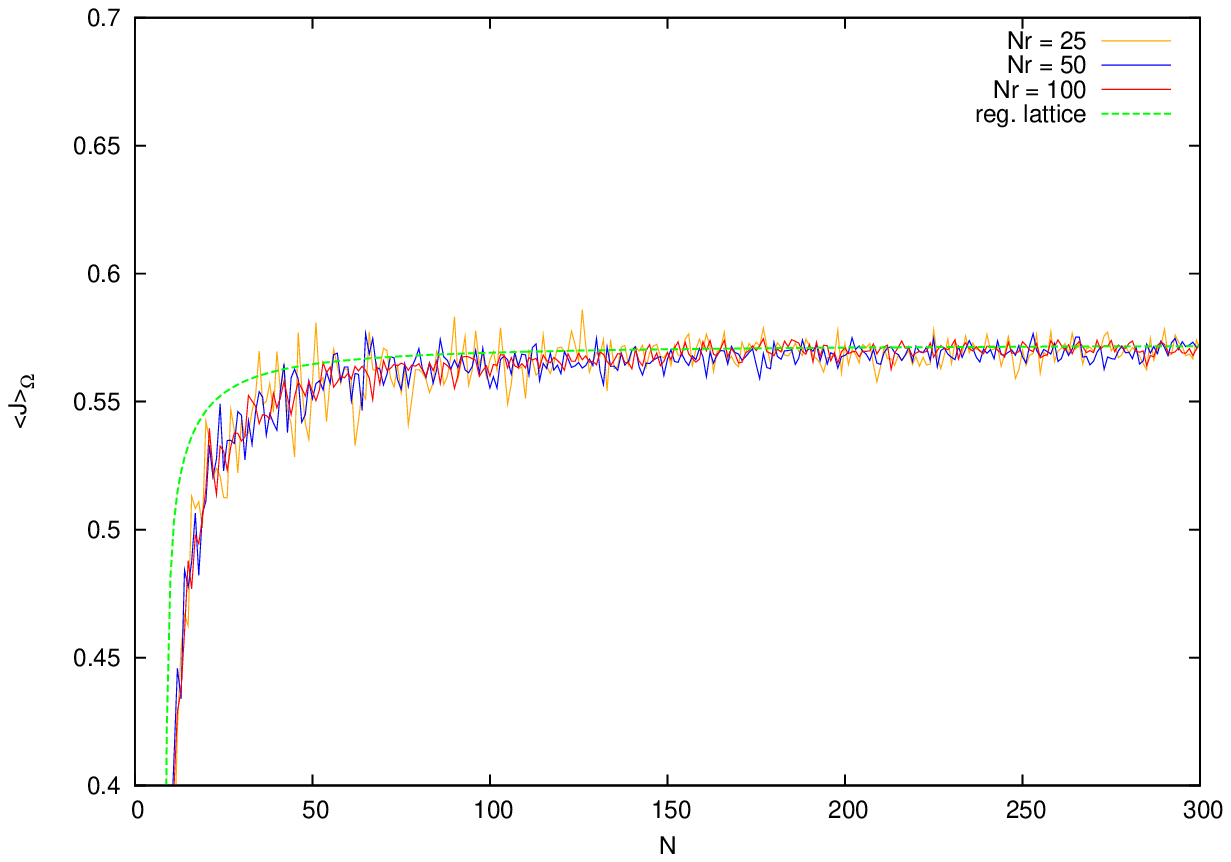}
   \hspace{1mm}
   \includegraphics[width=0.48\textwidth]{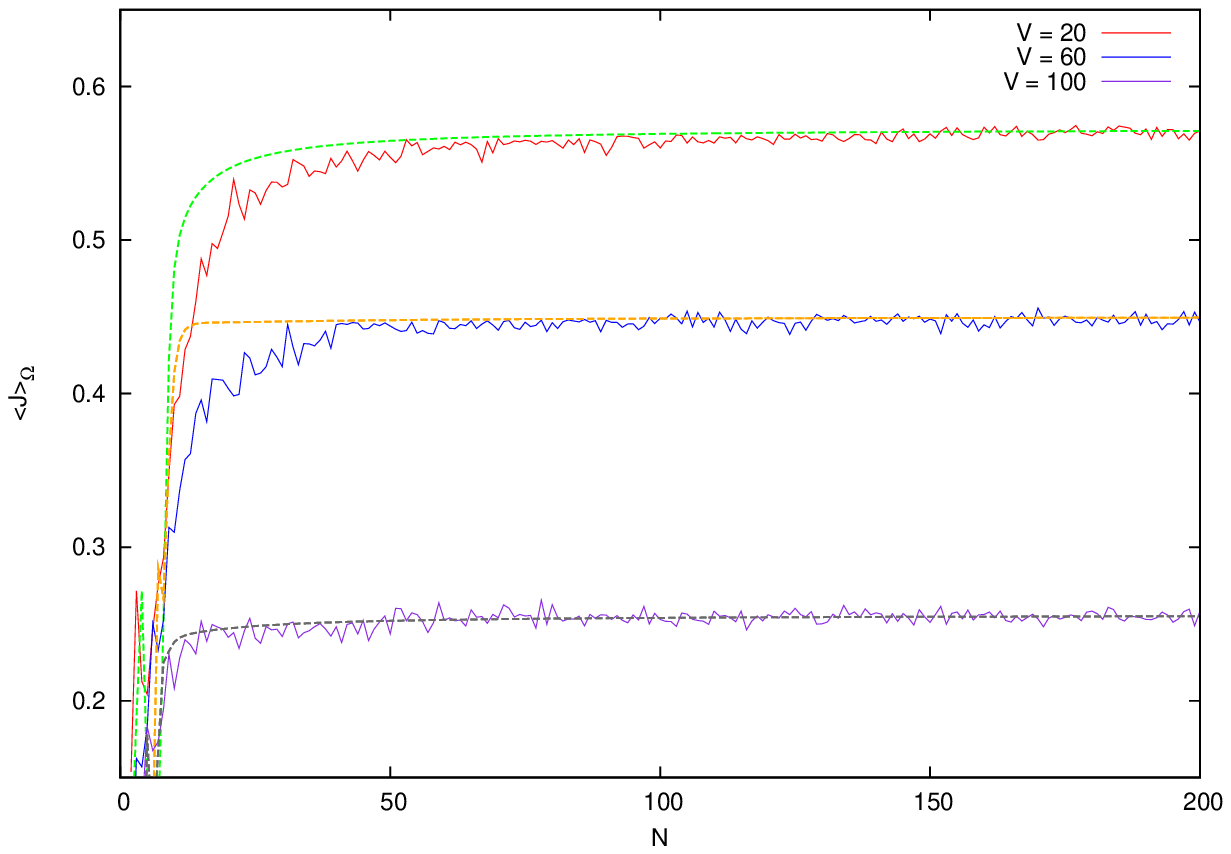}
   \caption{(Color online) \textit{Left panel}: Behavior of $\langle J \rangle_\Omega$ vs. $N$ for $V=20$, $T=300 K$, $E_v=1 eV$, $\beta = 0.1$ and $r=10^3$, in an ensemble of random uniform configurations for different values of $N_r$. \textit{Right panel}: Behavior of $\langle J \rangle_\Omega$ vs. $N$ for $N_r = 100$, $T=300 K$, $E_v=1 eV$, $r=10^3$ and for different values of $V$, in an ensemble of random uniform configurations. }\label{noise3}
\end{figure}

\begin{figure}
\centering
\includegraphics[width=0.48\textwidth]{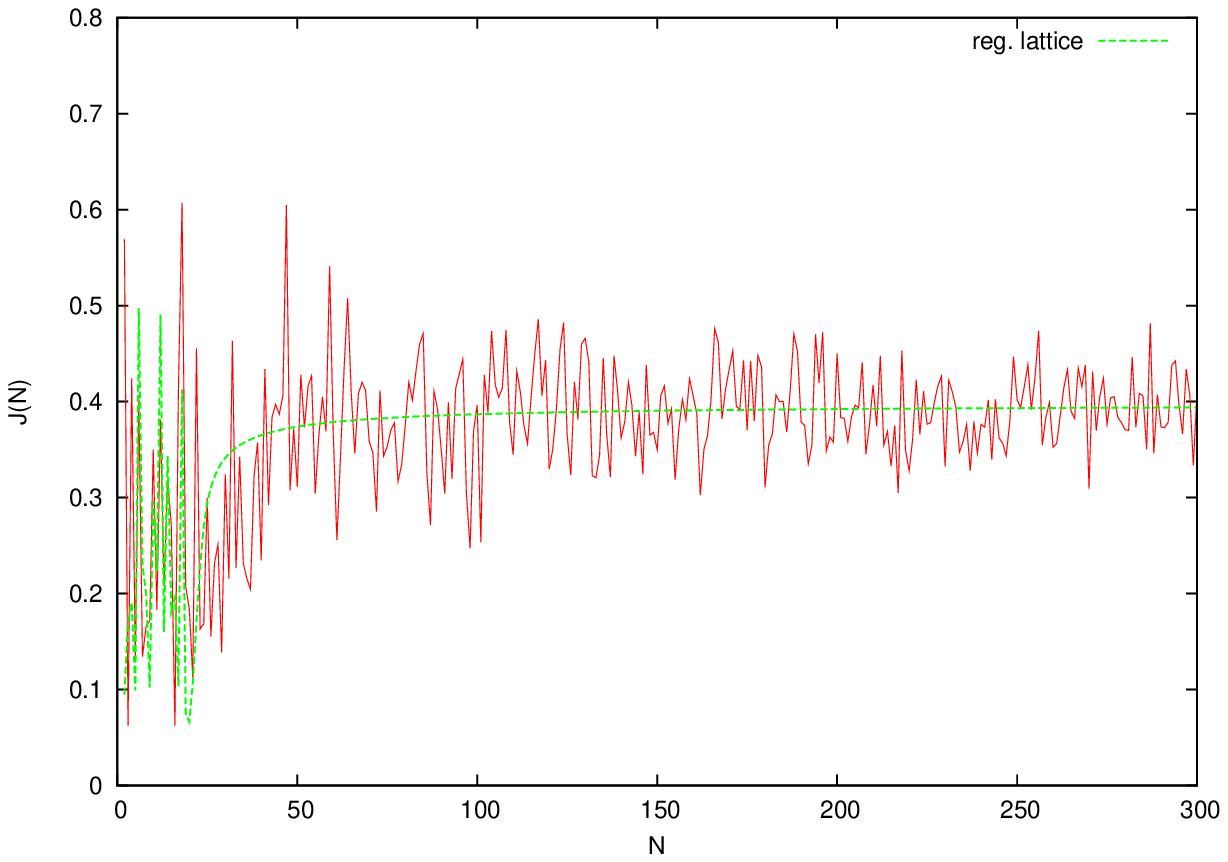}
\hspace{1mm}
\includegraphics[width=0.48\textwidth]{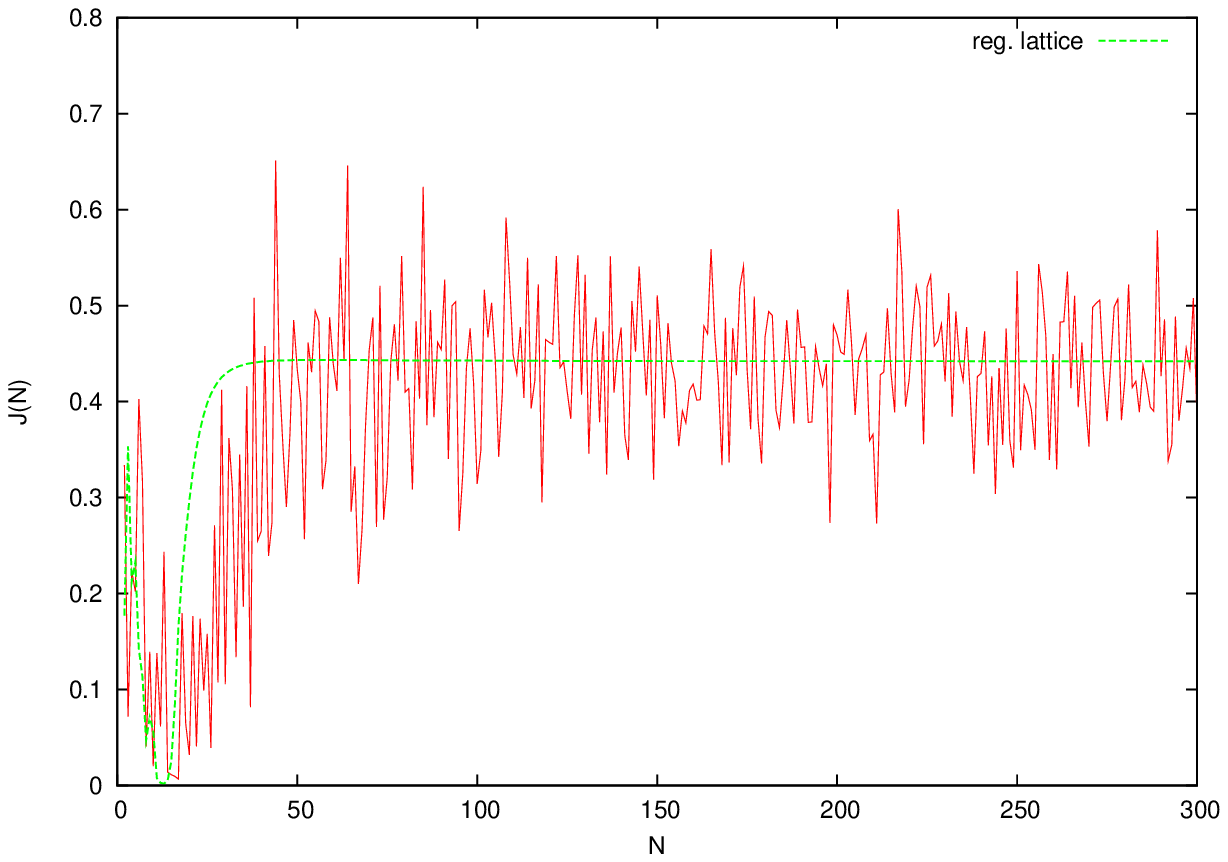}
\vspace{1mm}
\includegraphics[width=0.48\textwidth]{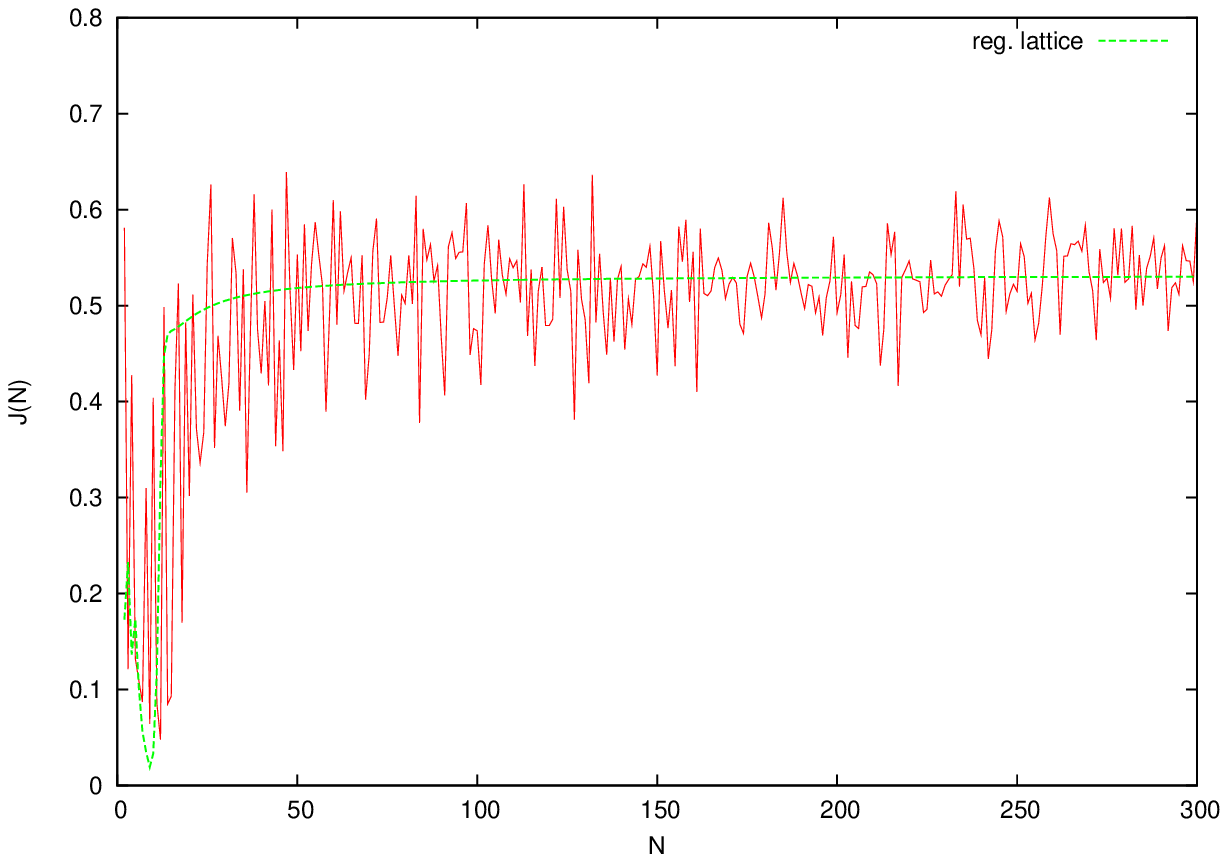}
\hspace{1mm}
\includegraphics[width=0.48\textwidth]{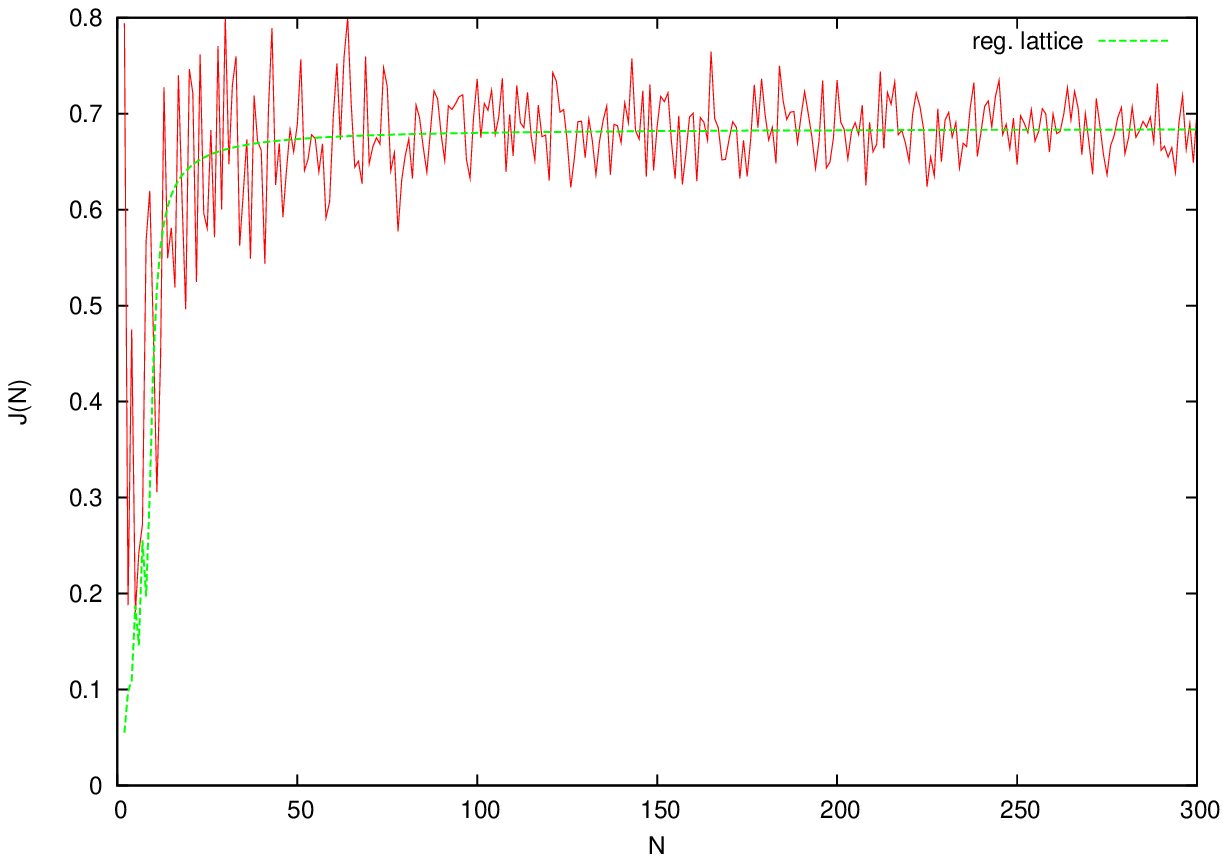}
\caption{(Color online) Behavior of $J(N,\Lambda_N,V;T)$ vs. $N$ with $V=20$, $T=300 K$, $E_v=1 eV$, $\beta = 0.1$ and, respectively, $r=10$ (top left), $r=10^2$ (top right), $r=10^3$ (bottom left) and $r=10^4$ (bottom right).}\label{noise4}
\end{figure}

\begin{figure}
\centering
\includegraphics[width=0.48\textwidth]{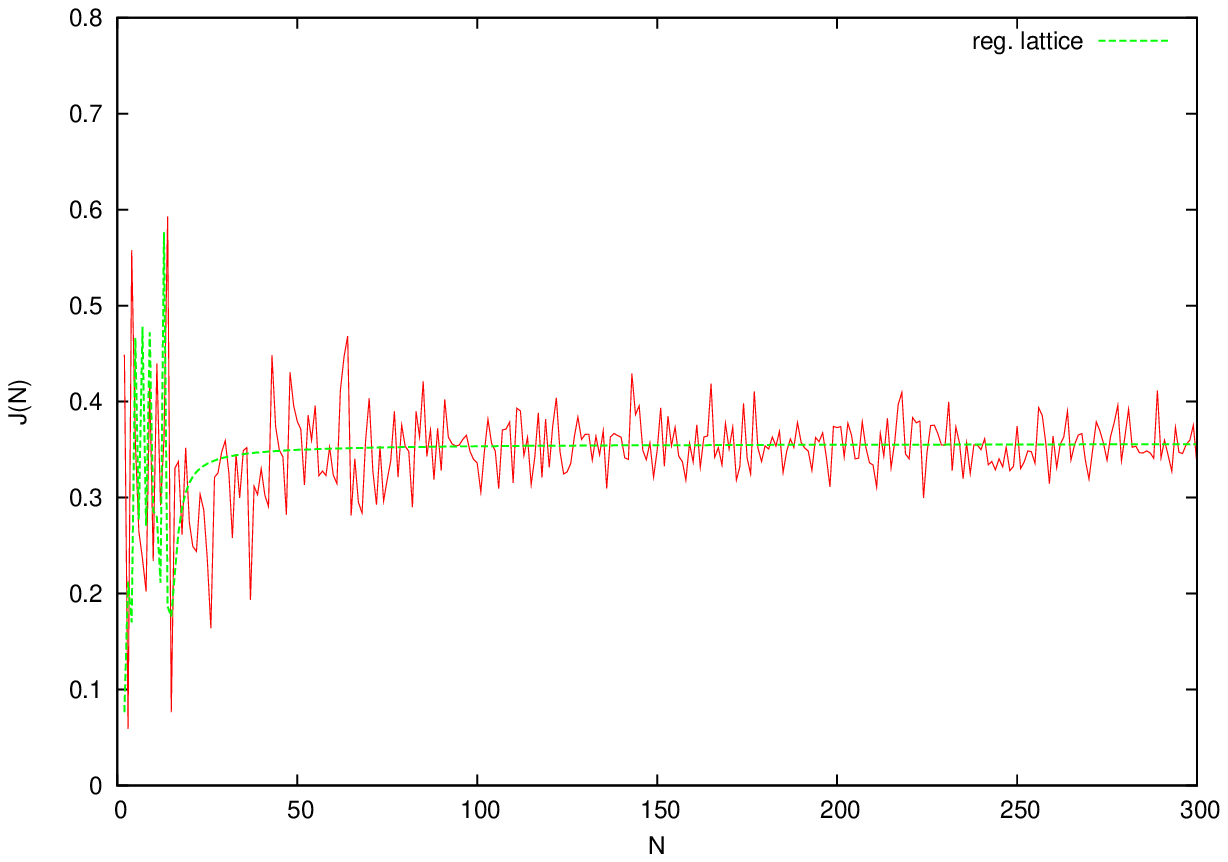}
\hspace{1mm}
\includegraphics[width=0.48\textwidth]{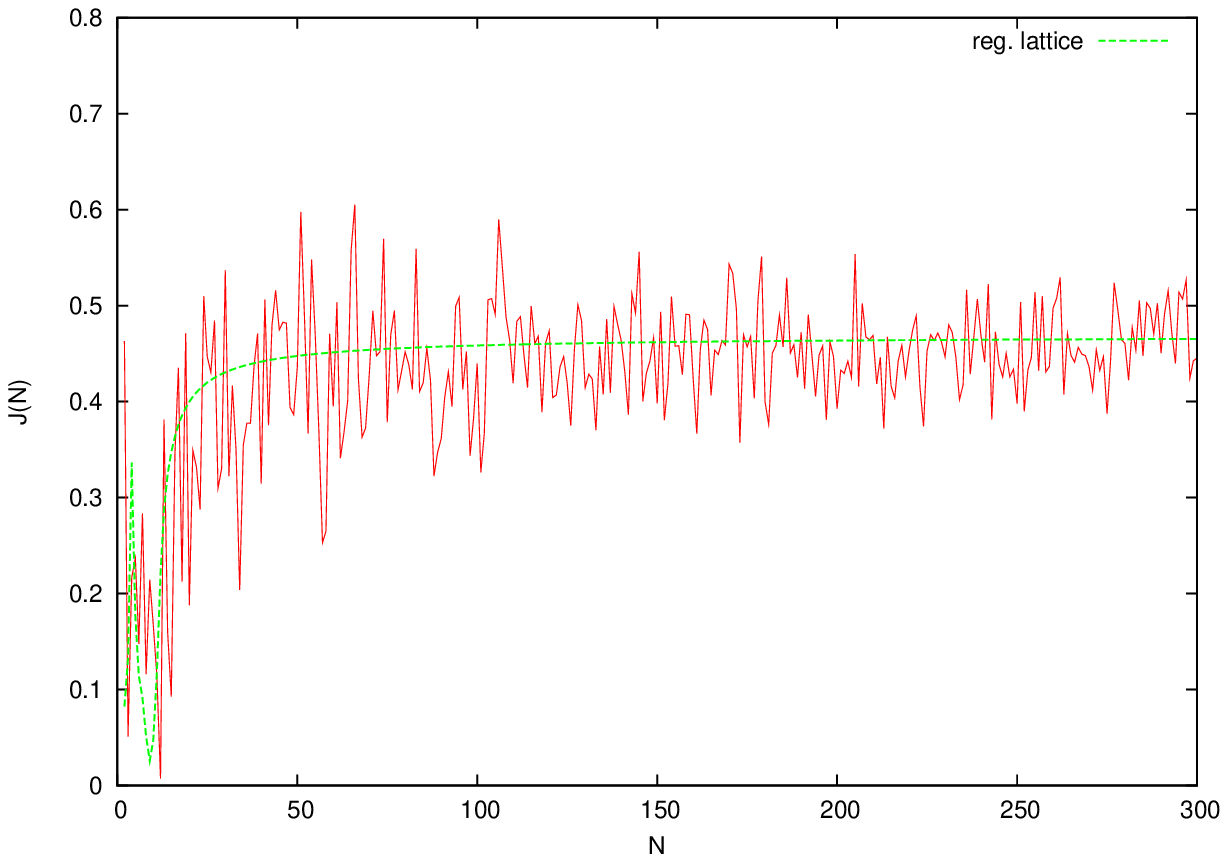}
\vspace{1mm}
\includegraphics[width=0.48\textwidth]{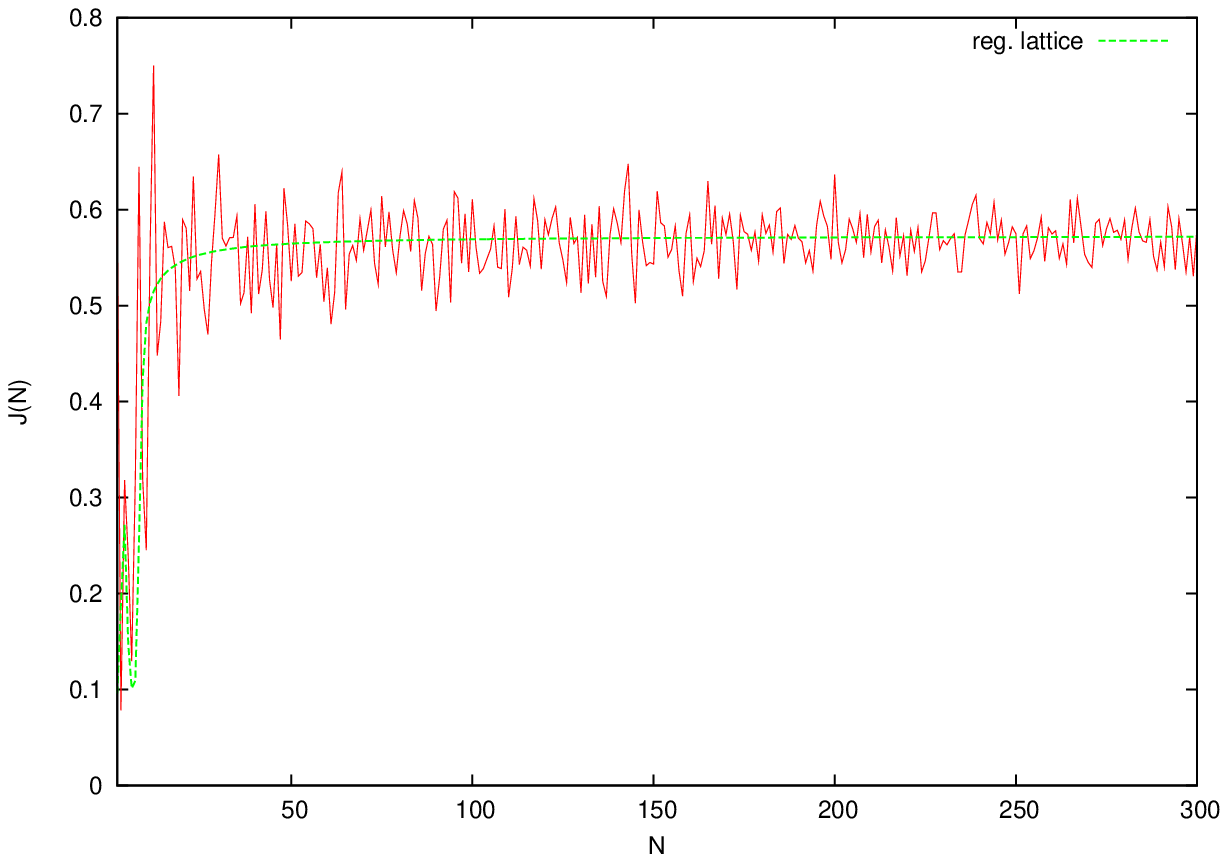}
\hspace{1mm}
\includegraphics[width=0.48\textwidth]{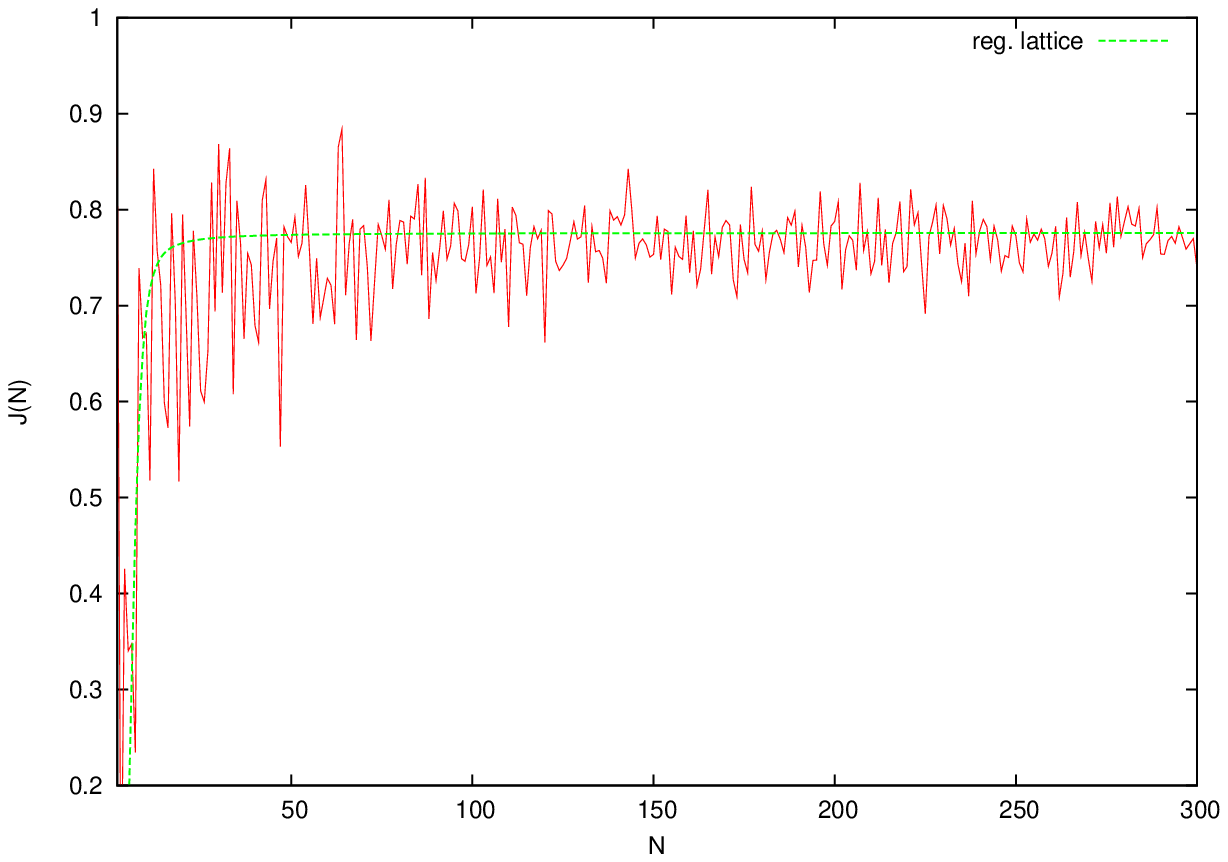}
\caption{(Color online) Behavior of $J(N,\Lambda_N,V;T)$ vs. $N$ with $V=20$, $T=300 K$, $E_v=2 eV$, $\beta = 0.1$ and, respectively, $r=10$ (top left), $r=10^2$ (top right), $r=10^3$ (bottom left) and $r=10^4$ (bottom right).}\label{noise5}
\end{figure}

Next, in the left panel of Fig. \ref{noise3}, we plotted the behavior of $\langle J \rangle_\Omega(N,V;T)$ with $N$ and we compared it with the value $J_B(N,V;T)$ pertaining to a regular lattice. The result is consistent with those of Figs. \ref{noise1} and \ref{noise2}, for it shows the regime corresponding to $N\gg 1$, where the curve of $\langle J\rangle_\Omega(N,V;T)$ varies very slowly with $N$, approaching the value $J_B(N,V;T)$. This convergence process is a collective effect, in that it is related to the delocalization of the wave function described by the Bloch waves theory \cite{mermin} for regular lattices. It is also worth pointing out that $J_B(N,V;T)$ resembles the value predicted by the Bloch theory only in the limit $N\gg1$, in which the model corresponds to a good approximation of the regular infinite lattice. 
The right panel of Fig. \ref{noise3} shows the behavior of $\langle J \rangle_\Omega(N,V;T)$ for different values of $V$, and highlights the dependence of $N_{macro}$ on $V$. In particular, the figure shows that taking $N> N_{macro}\approx 150,100,50$ for, respectively, $V=20,60,100$, allows one to reach a good accuracy even with small samples $\Omega$.\\
Moreover, we verified that the energies of the incoming wave, at $T = 300 K$, lie in the conducting band of the infinite periodic chain of barriers, which implies that $J_B(N,V;T)>0$. 
Furthermore, the plots in Figs. \ref{noise4} and \ref{noise5} show the behavior of $J(N,\Lambda_N,V;T)$ for different values of $E_v$ and $r$. At fixed $E_v$, we see that varying $r$ yields an increase of the transmission coefficient. Similarly, increasing $E_v$ results, in the observed region of the parameter space, in a slight increase of the transmission coefficient.

\begin{figure}
   \centering 
   \includegraphics[width=0.48\textwidth]{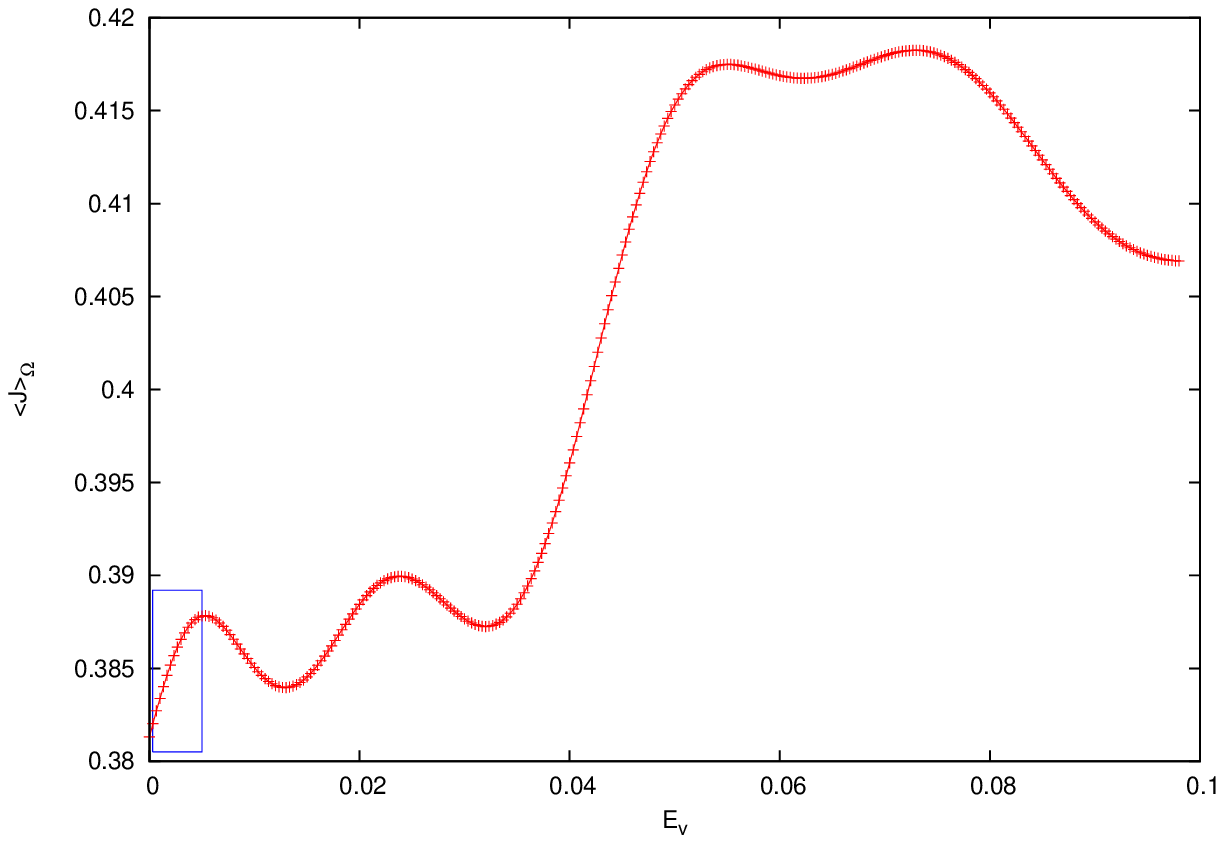}
   \hspace{1mm}
   \includegraphics[width=0.48\textwidth]{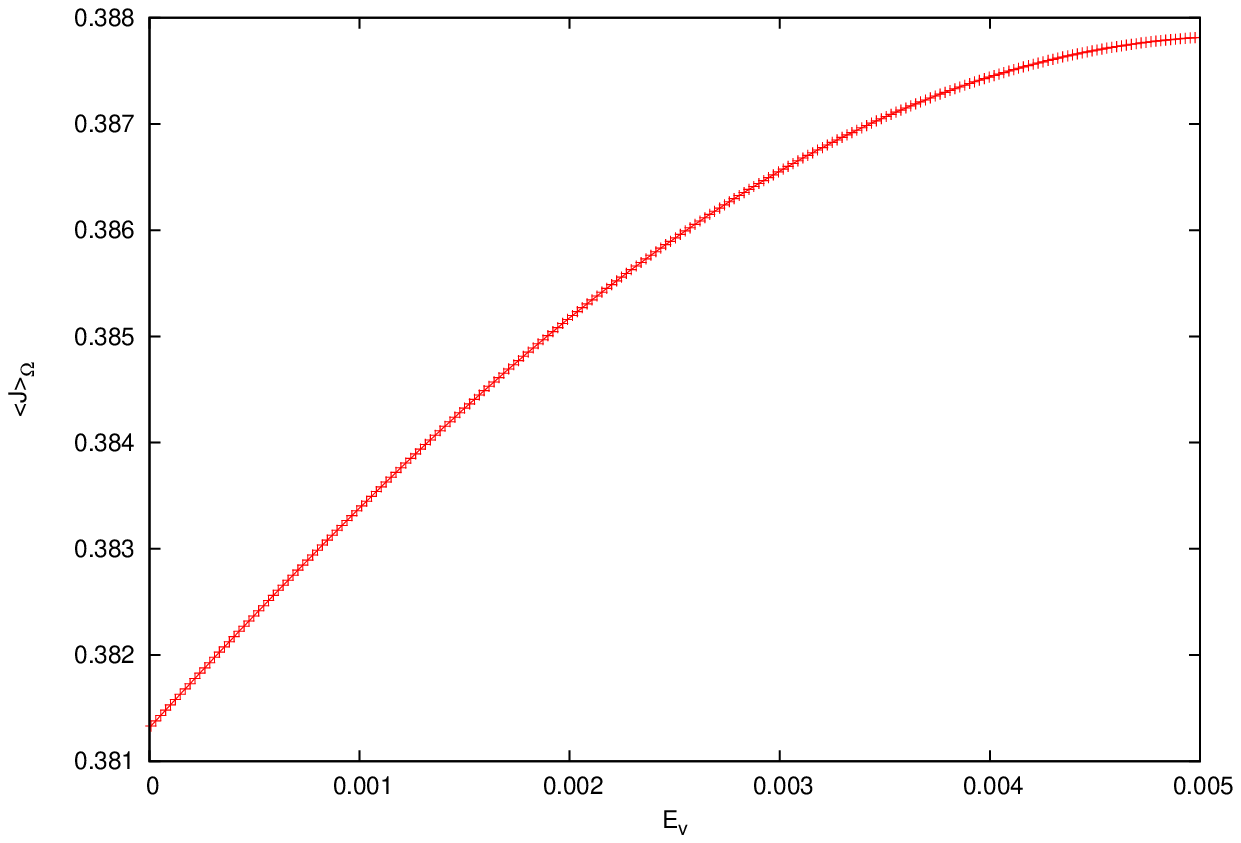}
   \caption{(Color online) \textit{Left panel}: Behavior of $\langle J \rangle_\Omega$ vs. $E_v$ for $N=300$, $V=20$, $T=300 K$, $\beta = 0.1$ and $r=10^2$, in an ensemble of $N_r=10^5$ random uniform configurations. \textit{Right panel}: Magnification of the framed part of the left panel.}\label{field}
\end{figure}

Figure \ref{field}, instead, shows the behavior of $\langle J \rangle_\Omega$, averaged over an ensemble of $N_r=10^5$ random uniform configurations, for different values $E_v$. It is worthwhile noticing that, in the limit of vanishing external fields, the value of the transmission coefficient is sensibly close to the value reported in \cite{ColRon}, referring to an equilibrium model.
A few comments can be drawn, here, also based on the comparison with those results discussed in Ref. \cite{ColRon}.
In the first place, the absence of localization can be traced back to the fixed finite amount of insulating material, which we have even in the $N\rightarrow\infty$ limit, because $L$ is fixed. As a consequence, incoming waves may, at most, be damped by a finite factor, except, perhaps, for a negligible set of energies which we have not observed. This distinguishes our model from the tight-binding model, which is more extensively investigated in the specialized literature, and also prevents the application of the Furstenberg's theorem \cite{Vulp}. Indeed, introducing the $(N+1)$-th barrier in one of our system realizations produces a rearrangement of the previous $N$ barriers. Mathematically, this means that the product of the first $N$ random matrices is replaced by a new product. Differently, the case of ergodic-like theorems, such as Furstenberg's theorem, applies to products of $N$ random matrices which do not change when they are multiplied by the $(N+1)$-th matrix. \\
The decrease of the size of fluctuations with $N$, which will be explored in more detailed below, can be regarded as a phenomenon of self-averaging of the observable $J$ \cite{Vulp}. In particular, our results, further supported by the analysis of the PDF of the transmission coefficient, Figs. \ref{prob} and \ref{ratefunct0} below, show that the random values $J(N,\Lambda_N,V;T)$ converge in probability to $J_B$ in the $N\rightarrow \infty$ limit. As shown in Fig. \ref{ratefunct0}, given a sample $\Omega$ of uniform realizations, $J_B(N,V;T)$ corresponds to the most probable value of the random variable $J$ in the sample, which, when $N$ grows, tends also to the mean $\langle J\rangle_\Omega(N,V;T)$.
Let us now investigate, more accurately, the structure of the fluctuations, in the sample $\Omega$ of $N_r$ random uniform configurations at temperature $T=300 K$.
Denote by $\rho_N(J)$ the probability density pertaining to the random value $J$.
The numerical results presented so far on the relation between $\langle J \rangle_\Omega(N,V;T)$ and $J_B(N,V;T)$, as well as on the decrease of the fluctuations size with growing $N$, indicate that $\rho_N(J)$ peaks more and more around the reference value $J_B(N,V;T)$. 
To show this more clearly, we numerically calculated the quantity $\rho_N(J)$ and we plotted in Fig. \ref{prob} the resulting curves for different values of $N$ and for $V=20$. The maxima of the PDF in Fig. \ref{prob} are approximately located at $J=J_B(N,V;T)$, cf. also Fig. \ref{ratefunct0}, and tend, for large $N$, to the mean value $\langle J \rangle_\Omega (N,V;T)$.
One further realizes that $\rho_N(J)$ obeys a sort of large deviation principle. By this we mean that the limit
\be
\zeta_N(J)\equiv\frac{-\log{\rho_N(J)}}{N}\xrightarrow{\scriptscriptstyle N\to\infty} \zeta(J) \label{largedev}
\ee
exists for the collection of $J$ values concerning the sample $\Omega$ of different realizations.
Figure \ref{ratefunct0} shows, for the range in which we have good statistics, that $\zeta$ is apparently smooth and strictly convex like a normal large deviation functional. However, it is worth pointing out, again, that the $N\rightarrow \infty$ limit is not achieved in the standard fashion of products of random matrices. Moreover, even the observable $J$ is not of the usual kind discussed in large deviation theory, in that it is not given by a sum of i.i.d. random variables, being it related, in general, to the random variables $\lambda_1,...,\lambda_N$ in a highly nonlinear fashion.

\begin{figure}
\begin{center}
  \includegraphics[width=0.7\textwidth]{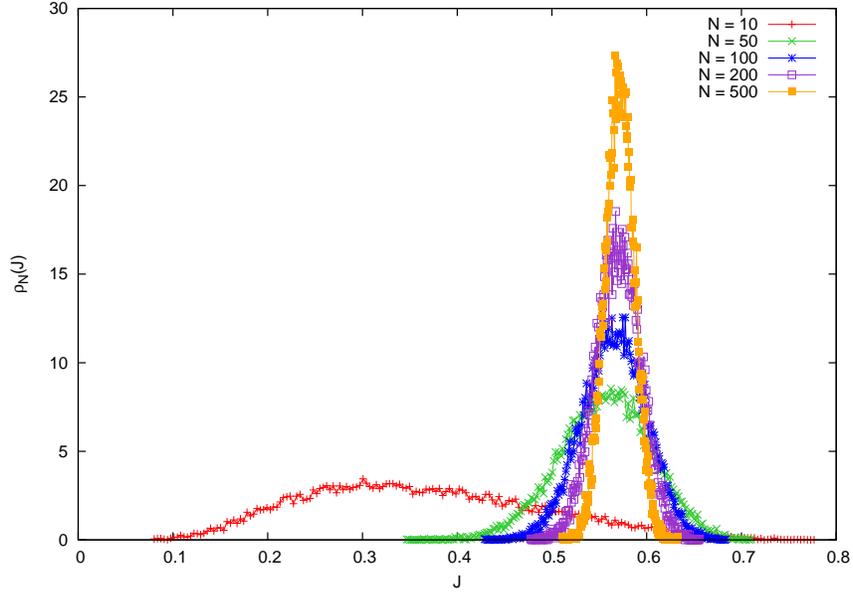}\\
  \caption{Probability densities $\rho_N(J)$ for different values of $N$ and for $V=20$, $T=300 K$, $E_v=1 eV$, $\beta = 0.1$ and $r=10^3$.}\label{prob}
   \end{center}
\end{figure}

\begin{figure}
\begin{center}
  \includegraphics[width=0.7\textwidth]{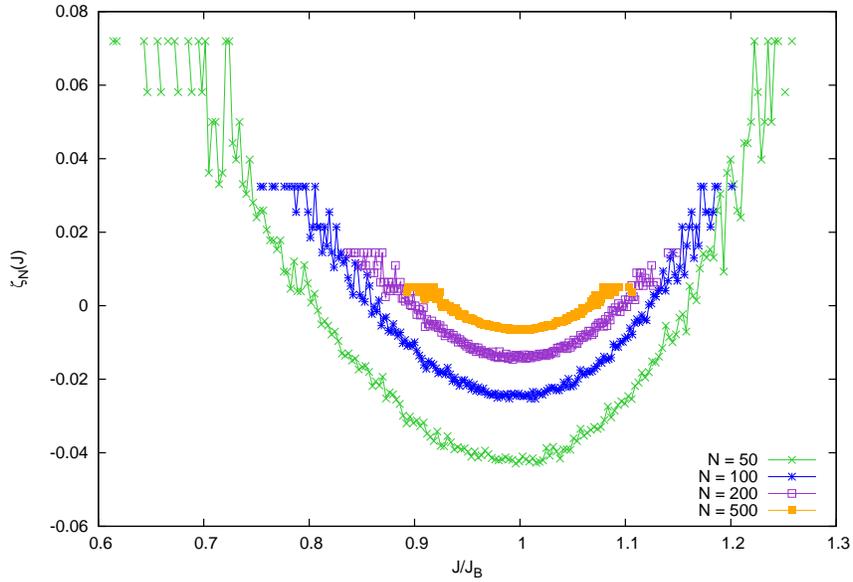}\\
  \caption{Rate functional $\zeta_N$ associated to $\rho_N(J)$ for different values of $N$ and for $V=20$, $V=20$, $T=300 K$, $E_v=1 eV$, $\beta = 0.1$ and $r=10^3$. The curves $\zeta_N$ move downwards for growing $N$, so that, as expected, in the $N\rightarrow \infty$ limit, $\zeta(J)$ intersects the horizontal axis only in $J=J_B$.}
\label{ratefunct0}
   \end{center}
\end{figure}

\begin{figure}
\begin{center}
  \includegraphics[width=0.7\textwidth]{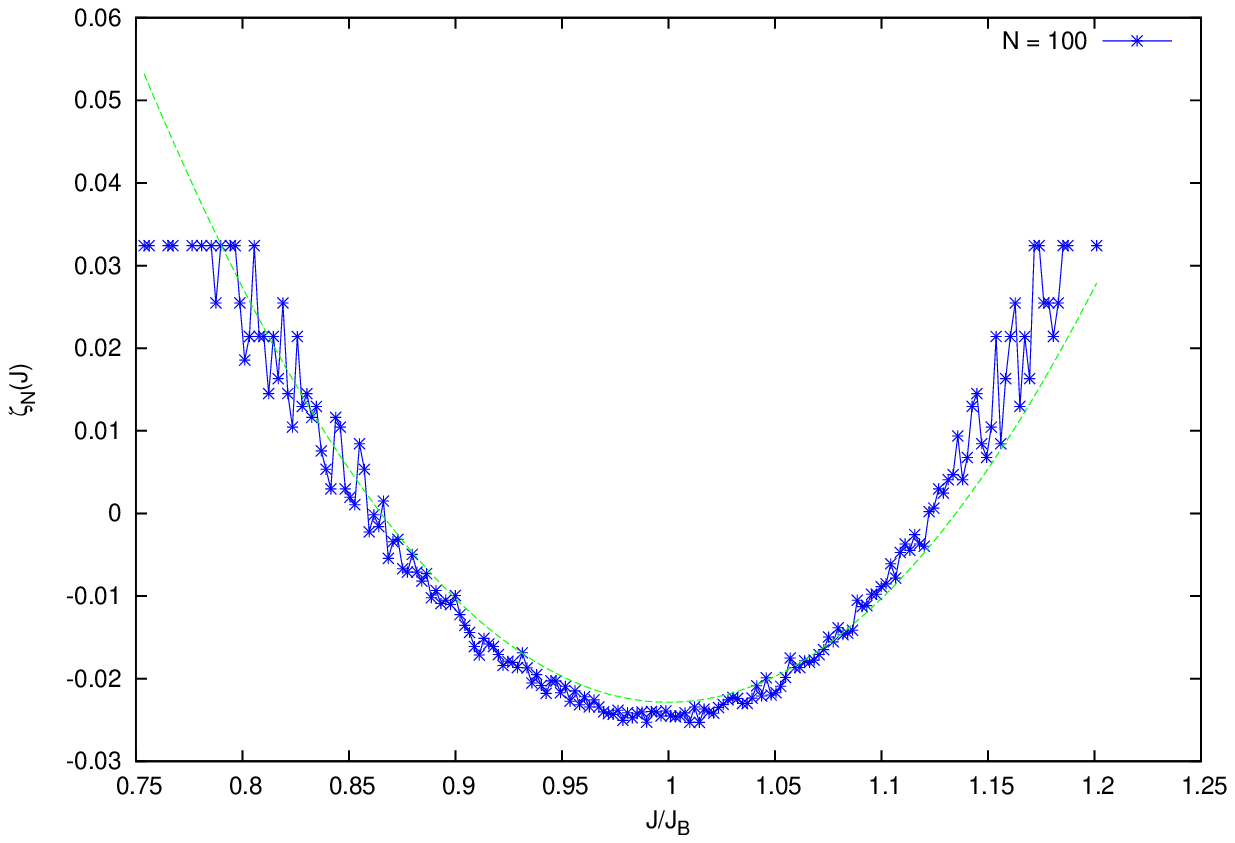}\\
  \caption{Rate functional $\zeta_N(J)$ associated to $\rho_N(J)$. \textit{Blue points}: results of the numerical simulation with $N=100$ and the same set of parameters used in Fig. \ref{prob}. \textit{Green dashed line}: Fitting of numerical data with the parabola $a(J/J_B-1)^2+b$, with parameters $a=1.25591 \pm 0.2575\cdot 10^{-1}$ and $b=-2.2858\cdot10^{-2}\pm 5.36\cdot10^{-4}$.}\label{EQclt}
   \end{center}
\end{figure}

It is also interesting to note that the locus of the minima, i.e. of highest probability density, of the curves $\zeta_N(J)$ is represented by the locus of the values $J_B(N,V;T)$, Fig. \ref{ratefunct0}.
Moreover, Fig. \ref{EQclt} reveals that, for small deviations from $J_B(N,V;T)$, $\zeta_N(J)$ is quadratic, as expected where the central limit theorem applies. 
\begin{figure}
   \centering
\includegraphics[width=8.5cm]{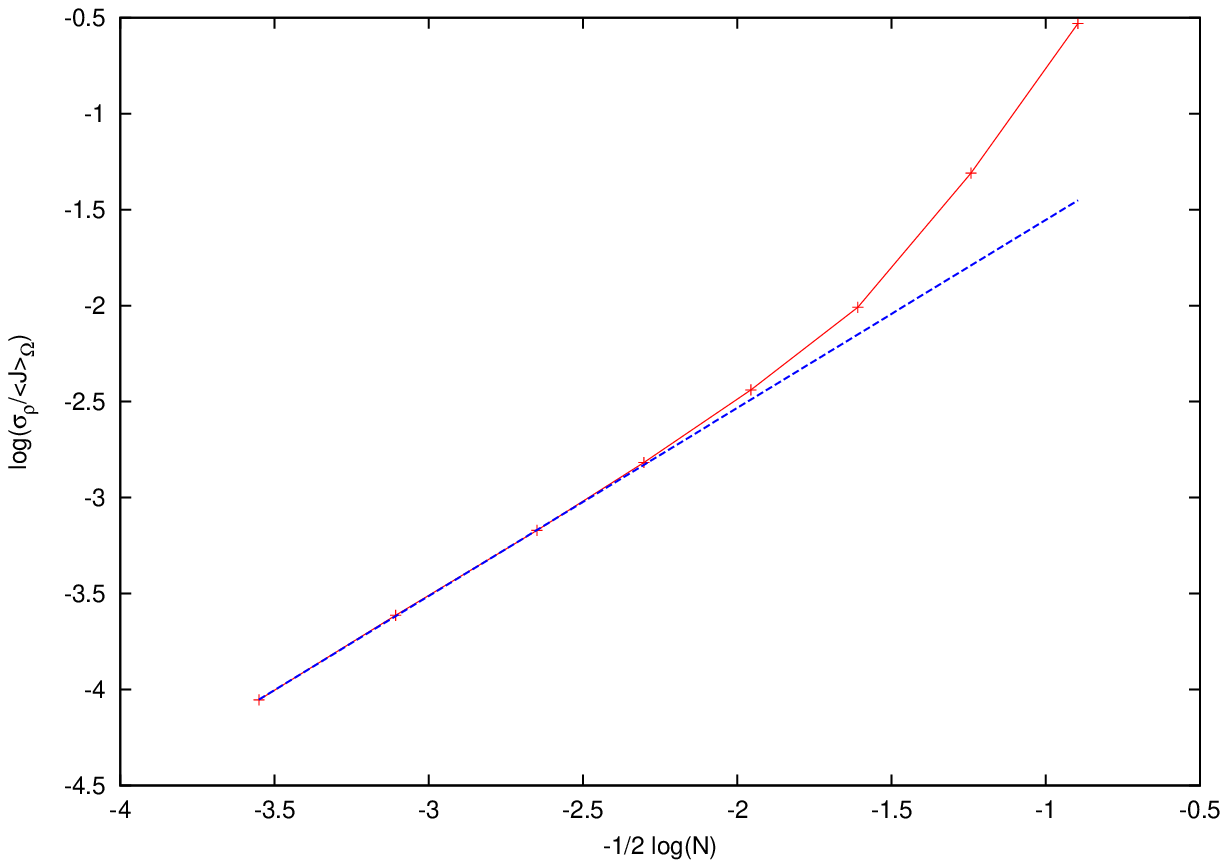}
\vspace{2mm}
\includegraphics[width=8.5cm]{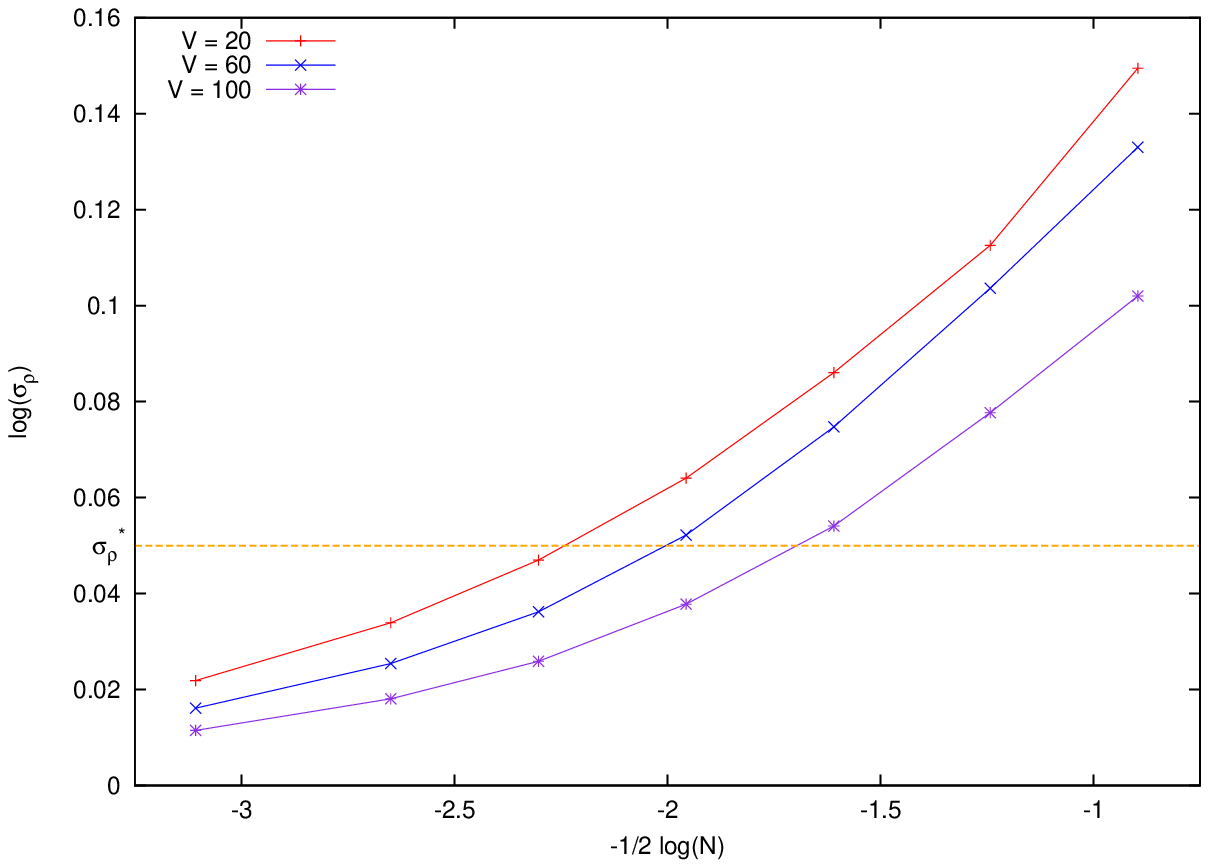}
   \caption{\textit{Top panel}: Fit of $\sigma_\rho/\langle J\rangle_\Omega$ (red curve), with the same set of parameters of Fig. \ref{prob}, with the curve $f(N)=a\times N^{-\frac{1}{2}}$, with $\log(a)=-5.13922\times 10^{-1} \pm 4.605 \times 10^{-3}$. The plot confirms the asymptotic behavior of $\sigma_\rho/\langle J\rangle_\Omega$, given by eq. (\ref{thermlim2}). \textit{Bottom panel}: Behavior of $\sigma_\rho(N,V)$  vs. $N$, for $V=\{20,60,100\}$. The thick orange dashed line denotes the value $\sigma_\rho^*$ below which we may conventially assume that the size of fluctuations is practically negligible.}\label{EQvariance}
\end{figure} 

From the validity of a principle of large deviations and of the central limit theorem, one expects the following asymptotic behavior for the fluctuations of the transmission coefficient:

\be
\frac{\sigma_\rho(N,V)}{\langle J\rangle_\Omega}\sim \left(\sqrt{N}\right)^{-1} \quad , \quad  \hbox{for $N\gg 1$}. \label{thermlim2}
\ee
We numerically checked that this is indeed the case by evaluating the ratio in eq. (\ref{thermlim2}) for $V=20$, cf. the top panel of Fig. \ref{EQvariance}. The bottom panel of Fig. \ref{EQvariance} further illustrates the decrease of $N_{meso}$ with $V$: taking $N_{meso}$ such that $\sigma_\rho(N,V)<\sigma_{\rho}^*=5.0\times 10^{-2}$ for $N>N_{meso}$, where $\sigma_{\rho}^*$ is considered small, $N_{meso}(V)$ is found to rapidly decrease with $V$: we obtain $N_{meso}\approx 90,55,30$ for, respectively, $V=20,60,100$. Then, $N_{meso}$ must tend to $0$ when the barrier height grows, because the transmission coefficient vanishes in this limit for any configuration $\Lambda_N$, hence the fluctuations are also suppressed. 

\section{Conclusions}

In this work we investigated a quantum mechanical model under nonequilibrium conditions, and focused on the role played by the disorder on the transmission coefficient $J$. Our numerical investigation reveals the existence of appropriate mesoscopic and macroscopic scales, respectively denoted by $N_{meso}$ and $N_{macro}$, which are not as widely separated as in thermodynamic systems.
The novelty of our approach stems, first, from the introduction of a thermal average of the transmission coefficient over an equilibrium distribution of energies guaranteed by the presence of an external thermostat at a given temperature $T$. Furthermore, we also proposed a novel approach to deal with the ``thermodynamic limit'' of the model: we prescribed a fixed, macroscopic, length for the system, so that the increase of the number of barriers does not yield the divergence of the overall length, rather it results in a more and more refined partition of barriers and wells.
The novel route proposed in this work leads, in general, to different results with respect to the standard thermodynamic limit discussed in the literature, and also makes some classical results, e.g. the Furstenberg's theorem, not applicable. Our approach might, hence, open a new line of investigation in the theory of disordered systems and could also allow to shed new light on the transition from the microscopic to the macroscopic scales.
Our numerical results suggest that, in presence of off-diagonal disorder, the wave function is delocalized over all the system length, thus no localization effect, of the like typically occurring with systems perturbed with diagonal disorder \cite{Izrailev}, occurs. Moreover, the disorder, at the microscopic level, induces an irregular behavior of the transmission coefficient $J$. The variable $J$ is self-averaging for growing $N$, and peaks over the most probable value of $J$. Interestingly, this value is $J_B$, which corresponds to a microscopic ordered array of barriers and conducting regions. Moreover, in the $N\gg 1$ limit, large deviations from the value $J_B$ are possible, and the structure of these fluctuations is governed by the rate functional $\zeta(J)$, which has been numerically determined. It is worth emphasizing that, at variance with the standard derivation of large deviation principles for the sum of independent and identically distributed random variables, our results hold for a random variable $J$ which is related, in general, in a highly nonlinear way to the random widths of the single barriers. Finally, this work suitably extends to a given nonequilibrium regime the results reported in \cite{ColRon}, obtained in absence of the external field.

\vskip 10pt
\noindent
{\bf Acknowledgments}

\vskip 5pt
M.C. wishes to thank Giuseppe Luca Celardo, Alberto Rosso, Alain Comtet and Christophe Texier for useful discussions.\\
L.R. gratefully acknowledges financial support from the European Research Council under the European
Community's Seventh Framework Programme (FP7/2007-2013) / ERC grant
agreement n 202680.  The EC is not liable for any use that can be made
on the information contained herein.


\vskip 10pt


\begin{thebibliography}{21}

\bibitem{dgm}
S. de~Groot, P. Mazur,\\ {\em Non equilibrium thermodynamics} (Dover, 1984).

\bibitem{Liboff}
R. Liboff,\\ {\em Kinetic Theory Classical, Quantum and Relativistic Descriptions} (Springer-Verlag, New York, 2003).

\bibitem{GibRon}  C. Giberti, L. Rondoni,\\ 
Anomalies and absence of local equilibrium, and universality, in one-dimensional particle systems, \\
\textit{Phys. Rev. E} \textbf{83}, 1 (2011).

\bibitem{matt2} M. Colangeli, I.V.Karlin, M. Kr\"{o}ger,\\
Hyperbolicity of exact hydrodynamics for three-dimensional Grad equations,
\textit{Phys. Rev. E} {\bf 76}, 022201 (2007).

 \bibitem{matt3}
I.V. Karlin, M. Colangeli, M. Kr\"oger, \\
Exact Linear Hydrodynamics from the Boltzmann Equation,
\textit{Phys. Rev. Lett.} {\bf 100}, 214503 (2008).

\bibitem{matt4} M. Colangeli, M. Kr\"{o}ger, H.C. \"{O}ttinger,\\
Boltzmann Equation and hydrodynamic fluctuations,
\textit{Phys. Rev. E} {\bf 80}, 051202 (2009).

\bibitem{touch} H. Touchette,\\
The large deviation approach to statistical mechanics,
\textit{Phys. Rep.} {\bf 478}, 1 (2009).

\bibitem{maes} M. Colangeli, C. Maes, B. Wynants,\\
A meaningful expansion  around detailed balance,\\
{\em J. Phys. A: Math. Theor.} \textbf{44} 095001 (2011).

\bibitem{colirr} M. Colangeli, L. Rondoni,\\
Equilibrium, fluctuation relations and transport for irreversible deterministic
dynamics,\\
{\em Physica D: Nonlinear Phenomena} {\bf 241}, 681 (2012).

\bibitem{BPRV} U. Marini Bettolo Marconi, A. Puglisi, L. Rondoni, A. Vulpiani,\\
Fluctuation-Dissipation: Response Theory in Statistical Physics,\\
\textit{Phys. Rep.} \textbf{461}, 111 (2008).

\bibitem{Vulp}
A. Crisanti, G. Paladin, A. Vulpiani, \\
{\em Product of random Matrices in Statistical Physics} (Springer-Verlag, Berlin, 1993).

\bibitem{Ander}
P. W. Anderson,\\ 
Absence of diffusion in certain random lattices,\\
\textit{Phys. Rev.} \textbf{109} 1492 (1958).

\bibitem{TC} G. Theodorou, M. H. Cohen,\\
Extended states in a one-dimensional system with off-diagonal disorder,\\ 
{\em Phys. Rev. B} \textbf{13}, 10 (1976).

\bibitem{SE} C. M. Soukoulis, E. N. Economou,\\ 
Off-diagonal disorder in one-dimensional systems,\\ 
{\em Phys. Rev. B} \textbf{24}, 10 (1981).

\bibitem{Izrailev} F. M. Izrailev, A. A. Krokhin, N. M. Makarov,\\
Anomalous localization in low-dimensional systems with correlated disorder,\\
{\em Phys. Rep.} \textbf{512}, 125 (2012).

\bibitem{ColRon} M. Colangeli, L. Rondoni\\
Fluctuations in quantum one-dimensional thermostatted systems with off-diagonal disorder,\\
{\em J. Stat. Mech.} P02009 (2013).

\bibitem{harris}
P. Harrison,\\
\textit{Quantum wires, wells and dots} (John Wiley and sons, Ltd, 2005).

\bibitem{mermin}
N. W. Ashcroft, N. D. Mermin,\\
\textit{Solid State Physics} (Brooks/Cole, Thomson Learning, Cornell, 1976).

 
\end{thebibliography}
\end{document}